\documentclass[nofootinbib,aps,a4paper,10pt,superscriptaddress,
onecolumn,showkeys,times,eqsecnum]{revtex4-2}
\usepackage{amsmath}
\usepackage{amsmath,tabstackengine}
\usepackage{amsfonts}
\usepackage{tabularx}
\usepackage{framed}
\usepackage{nccmath} 
\usepackage[utf8]{inputenc}
 \pdfoutput=1
\usepackage{amsmath}
\usepackage{colortbl}
\usepackage{array}       
\usepackage{tabularx}     
\usepackage{lipsum} 
\usepackage{mathtools} 
\usepackage{array}
\usepackage{multirow}
\usepackage{makecell}
\usepackage{booktabs}
\usepackage{graphicx}
\usepackage{multirow}
\usepackage{color}
\usepackage{braket}
\usepackage{dcolumn}
\usepackage{siunitx}
\usepackage{bm,url}
\usepackage{colortbl}
\usepackage[linktocpage]{hyperref}
\usepackage{subfigure}
\usepackage{amsfonts}

\begin{document}

\title{Thick branes and fermion localization in five-dimensional 
 $f(T,T_G)$ gravity}

\author{A. R. P. Moreira}
\email{allan.moreira@fisica.ufc.br}
\affiliation{Research Center for Quantum Physics, Huzhou Normal University, Huzhou, 
313000, P. R. China.}
\affiliation{Secretaria da Educaç\~{a}o do Cear\'{a} (SEDUC), Coordenadoria 
Regional de Desenvolvimento da Educaç\~{a}o (CREDE 9),  Horizonte, Cear\'{a}, 
62880-384, Brazil.}
\author{F. M. Belchior}
 \email{belchior@fisica.ufc.br}
\affiliation{Departamento de Física, Universidade Federal da Paraíba, Centro de Ciências Exatas e da Natureza, 58051-970, João Pessoa, Paraíba, Brazil}
\author{Shi-Hai Dong}
 \email{dongsh2@yahoo.com}
\affiliation{Research Center for Quantum Physics, Huzhou Normal University, Huzhou, 
313000, P. R. China.}
\affiliation{Centro de Investigaci\'{o}n en Computaci\'{o}n, Instituto 
Polit\'{e}cnico Nacional, UPALM, CDMX 07700, M\'{e}xico}
\author{E. N. Saridakis}
\email{msaridak@noa.gr}
\affiliation{National Observatory of Athens, Lofos Nymfon, 11852 Athens, Greece}
\affiliation{CAS Key Laboratory for Researches in Galaxies and Cosmology, 
School 
of Astronomy and Space Science, University of Science and Technology of China, 
Hefei, Anhui 230026, China}
\affiliation{Departamento de Matem\'{a}ticas, Universidad Cat\'{o}lica del 
Norte, Avda. Angamos 0610, Casilla 1280 Antofagasta, Chile}

\begin{abstract}
We investigate thick-brane configurations in five-dimensional $f(T,T_G)$ 
modified teleparallel gravity. In five dimensions, the torsional Gauss-Bonnet 
invariant $T_G$ contributes dynamically, leading to genuinely new effects even 
at linear order. Within a warped geometry supported by a scalar field, we 
construct explicit solutions and show that the $T_G$ sector significantly 
modifies the brane structure. In particular, the coupling  parameter  
controls the deformation of the warp factor and energy density, allowing for 
the emergence of brane splitting and nontrivial internal structure. 
We further analyze the localization of spin-$1/2$ fermions via a Yukawa 
coupling. The system admits a normalizable chiral zero mode, while the opposite 
chirality remains delocalized. The massive Kaluza-Klein spectrum is strongly 
affected by the torsional Gauss-Bonnet term, which modifies the effective 
potentials and leads to the appearance of resonant quasi-localized states.
Our results show that $f(T,T_G)$ gravity provides a richer framework for 
braneworld models, where torsional higher-order corrections play a key role in 
shaping both geometry and field localization.
\end{abstract}


\maketitle

\section{Introduction}

The idea that our observable Universe may be embedded in a higher-dimensional 
spacetime has long provided a powerful framework for addressing fundamental 
problems in high-energy physics and cosmology. Originally motivated by the 
hierarchy problem and further developed within string-theoretic constructions, 
extra-dimensional models offer novel mechanisms for localizing matter fields, 
modifying gravitational interactions, and generating rich cosmological dynamics 
\cite{Kaluza:1921tu,Klein:1926tv}. In this context, braneworld scenarios have 
attracted considerable attention, as they allow for a consistent realization of 
four-dimensional physics on a hypersurface embedded in a higher-dimensional 
bulk  \cite{Arkani-Hamed:1998jmv,Antoniadis:1998ig,Sahni:2002dx, Shtanov:2002ek, 
Saridakis:2007wx,   Lobo:2007qi, Schmidt:2009sg,  
Abdujabbarov:2009az, Kudoh:2003xz, Majumdar:2005ba, Eiroa:2004gh, Setare:2008mb, 
Schmidt:2009sv, Lombriser:2009xg, Bento:2002np, Koyama:2007ih,  
Saridakis:2007cf, Schee:2008fc, Alam:2005pb, Lidsey:2002zw, Lazkoz:2006gp, 
Schmidt:2009yj, Calcagni:2004bh, Abdujabbarov:2017pfw, Tretyakov:2005en}.
Among the various realizations, thick-brane models, where the brane is 
described as a smooth domain-wall configuration supported by bulk 
fields, provide a particularly appealing and realistic framework 
\cite{Randall:1999vf,Randall:1999ee}. In contrast to thin-brane constructions, 
thick branes allow for a fully dynamical and regular treatment of the 
gravitational and matter sectors 
\cite{Gremm:1999pj,Bazeia:2008zx,Dzhunushaliev:2009va}. A central question in 
this setting is how the localization of gravity and matter fields is affected 
when the bulk gravitational dynamics deviates from General Relativity. 
Addressing this issue is essential for understanding the robustness of 
braneworld scenarios and for identifying possible observational signatures of 
extra dimensions 
\cite{Charmousis:2001hg,Arias:2002ew,Zhong:2022wlw,Peyravi:2022ubf,
Gordin:2023nsv,Tan:2020sys,Azizi:2025kwl, Deng:2025hfn,Liu:2009ve,
Liu:2009mga}.

On the  other hand, Teleparallel gravity offers an alternative geometrical 
formulation of gravitation, in which torsion, rather than curvature, encodes the 
gravitational interaction. The teleparallel equivalent of General Relativity 
(TEGR) reproduces Einstein's equations while being constructed from the torsion 
scalar $T$ associated with a curvatureless connection 
\cite{Aldrovandi:2013wha,Cai:2015emx}. This formulation is particularly 
suitable for modified gravity model building, since extensions of the form 
$f(T)$ prove to lead to interesting phenomenology
\cite{Chen:2010va, Bengochea:2010sg, Karami:2010bys,  
HamaniDaouda:2011iy, Meng:2011ne, Karami:2012fu, Tamanini:2012hg, 
Cardone:2012xq, Salako:2013gka, Nashed:2013bfa, Ong:2013qja, Otalora:2014aoa, 
Farrugia:2016xcw,  Cai:2018rzd, 
Ferraro:2018tpu,vandenHoogen:2023pjs, Jiang:2024otl, Zhao:2024uzq, 
Fenwick:2024jby, Jawad:2025mzv, SwagatMishra:2025wut, Landry:2025whg,
Manzoor:2026pyq,Bouhmadi-Lopez:2026dte}. 
The study of $f(T)$ gravity in braneworld scenarios has revealed a rich 
phenomenology, including modified brane structures, nontrivial localization 
properties, and the emergence of resonant spectra \cite{Menezes:2014bta, 
Yang:2012hu,Guo:2015qbt,Wang:2018jsw,Yang:2017evd,Guo:2018tpo,
Moreira:2021xfe,Moreira:2021vcf,Moreira:2021uod,Moreira:2023pes,
Moreira:2024zio}. However, these models remain limited in the sense that they do 
not incorporate higher-order geometric invariants, which are known to play a 
crucial role in higher-dimensional gravity.

A natural next step is therefore to extend the theory by including higher-order 
torsional invariants. In curvature-based gravity, the Gauss-Bonnet term 
represents the simplest such extension, possessing special properties: it is 
topological in four dimensions, while in higher dimensions it contributes 
nontrivially to the dynamics and yields second-order field equations as part of 
Lovelock gravity. Inspired by this structure, Kofinas and Saridakis introduced 
the teleparallel equivalent of the Gauss-Bonnet invariant $T_G$ 
\cite{Kofinas:2014owa}, constructed from torsion (or contortion) 
variables and related to the curvature Gauss-Bonnet scalar up to a total 
divergence.  
This development opened the way to $f(T,T_G)$ gravity 
\cite{Kadam:2024rwd,Kadam:2022daz,Azhar:2020coz,Sharif:2018sgg,Farrugia:2018gyz,
delaCruz-Dombriz:2017lvj}, which provides a genuinely new class of modified 
gravitational theories. In particular, these models differ both from 
curvature-based $f(R,G)$ theories and from purely torsional $f(T)$ models, while 
preserving a geometric structure closely related to Lovelock gravity 
\cite{Ilyas:2025pvh,Rehman:2025ikt,Moreira:2025vsc,Ilyas:2025fal,
Samaddar:2025dyj,Akbarieh:2025vau,Dimakis:2025wtc,Ilyas:2026qbi,Dubey:2025nuf}. 

In this context, braneworld scenarios in $f(T,T_G)$ gravity provide an ideal 
laboratory to investigate the interplay between extra dimensions, torsion-based 
modifications, and higher-order invariants. A crucial point, often 
overlooked, is that, in five dimensions, the torsional Gauss-Bonnet invariant 
$T_G$ is no longer a boundary term, but contributes dynamically to the field 
equations. This implies that even the simplest extension $f(T,T_G)=-T+\alpha 
T_G$   leads to qualitatively new features in the structure of thick branes, 
which are absent in both $f(T)$ models and curvature-based Gauss-Bonnet 
braneworlds.
In particular, the presence of the $T_G$ sector introduces additional 
derivative structures through $f_{T_G}$ and its derivatives along the extra 
dimension, affecting the warp factor, the scalar-field configuration, and the 
resulting effective potentials governing field localization. Consequently, one 
expects modifications not only in the background geometry, but also in the 
localization of matter fields and in the structure of the Kaluza-Klein 
spectrum, including the possible emergence and redistribution of resonant modes.

The aim of the present work is to construct and analyze thick-brane solutions 
in five-dimensional $f(T,T_G)$ gravity and to investigate the impact of the 
torsional Gauss-Bonnet sector on both the geometric and fermionic sectors. We 
focus on the minimal yet nontrivial model  in order to 
isolate and clearly identify the physical effects induced by the $T_G$ 
contribution.
Thus, we consider a warped geometry supported by a canonical scalar 
field and derive the full set of background equations governing the system. We 
then obtain explicit brane configurations and analyze how the parameter 
$\alpha$ controls the internal structure of the brane, including the emergence 
of brane splitting. Subsequently, we study the localization of spin-$1/2$ 
fermions via a Yukawa coupling, examining both the zero-mode sector and the 
massive Kaluza-Klein spectrum, with particular emphasis on the formation of 
resonant states.

The manuscript is organized as follows. In Sec.~\ref{s2} we review the 
essentials of teleparallel geometry, the torsion scalar $T$, and the 
teleparallel Gauss-Bonnet invariant $T_G$, and we present the $f(T,T_G)$ action 
and general field equations. Additionally, we introduce the warped brane 
ansatz, compute $T$ and $T_G$ explicitly, and derive the coupled ordinary 
differential equations for the background functions. In Sec.~\ref{s3} we discuss 
brane solutions for selected functional forms of $f(T,T_G)$ and potentials, 
including the parameter dependence and limiting cases. In Sec.~\ref{s4} we 
study localization of fermions, derive the localization conditions for the 
fermion zero mode, and analyze the massive Kaluza-Klein sector. In Sec.~\ref{s5}
we discuss the physical implications of the torsional Gauss-Bonnet 
term. 
Finally, Sec.~\ref{s6} summarizes our results and outlines 
possible extensions, including vector and scalar perturbations, fermion 
localization, and cosmological applications.

\section{$5$-dimensional $f(T,T_G)$ gravity and thick-brane system}\label{s2}

In this section  we   present the  $5$-dimensional ($5$D) extension of 
teleparallel gravity with Lagrangian $f(T,T_G)$, where $T$ is the torsion scalar 
and $T_G$ is the teleparallel equivalent of the Gauss-Bonnet invariant 
following \cite{Kofinas:2014owa}, and we derive the
reduced equations of motion for warped thick-brane geometries supported by a 
bulk scalar field.

\subsection{Teleparallel variables and invariants}
\label{subsec:teleparallel_invariants}

The fundamental field is the fünfbein (the five-dimensional analogue of the 
vielbein/tetrad) $e^{a}{}_{M}$ (with $a=0,1,2,3,4$ marking tangent indices, and 
$M=0,1,2,3,4$ spacetime indices), with determinant 
$e\equiv\det(e^{a}{}_{M})=\sqrt{-g}$. The metric in terms of fünfbein is written 
as
\begin{align}\label{eq1}
g_{MN}=\eta_{ab}\,e^{a}{}_{M}e^{b}{}_{N},
\end{align}
where $\eta_{ab}=\mathrm{diag}(-,+,+,+,+)$. We then employ the teleparallel 
(curvature-free) connection, and for concreteness, we work in the Weitzenb\"ock 
(``pure tetrad'') gauge  where the inertial spin connection vanishes. Such a 
connection is explicitly given by 
\begin{align}\label{eq2}
\widetilde{\Gamma}^{P}{}_{MN}=e_{a}{}^{P}\partial_{N}e^{a}{}_{M}.    
\end{align}
The torsion tensor is then
\begin{align}\label{eq3}
T^{P}{}_{MN}=\widetilde{\Gamma}^{P}{}_{NM}-\widetilde{\Gamma}^{P}{}_{MN},
\end{align}
and the torsion scalar $T$ is constructed from the quadratic contractions of 
torsion, namely
\begin{align}\label{eq4}
T=\frac14\,T_{PMN}T^{PMN}+\frac12\,T_{PMN}T^{NMP}-T_{P}T^{P},
\end{align}
where $T_{P}\equiv T^{M}{}_{MP}$ is the trace of torsion tensor. In this case, 
$T$ is the standard TEGR invariant as defined 
in~\cite{Kofinas:2014owa}. Besides, the teleparallel
equivalent of Gauss-Bonnet gravity is built from the torsion/contortion in such 
a way that it reproduces the usual curvature Gauss-Bonnet invariant up to a 
total divergence. To construct this invariant, one defines the contortion 
(tangent components) as
\begin{align}\label{eq5}
K^{a}{}_{bc}=\frac12\Big(T_{b}{}^{a}{}_{c}+T_{c}{}^{a}{}_{b}-T^{a}{}_{bc}\Big).
\end{align}

With these definitions at hand we can write
the teleparallel 
Gauss-Bonnet $D$-form in the differential-form language  as  
\cite{Kofinas:2014owa}
\begin{align}\label{eq6}
T_G&=\frac{1}{(D-4)!}\,\epsilon_{a_1\cdots a_D}
\Big(
K^{a_1}{}_{c}\wedge K^{ca_2}\wedge K^{a_3}{}_{d}\wedge K^{da_4}
-2\,K^{a_1a_2}\wedge K^{a_3}{}_{c}\wedge K^{cd}\wedge K_{d}{}^{a_4}
\nonumber\\&+2\,K^{a_1a_2}\wedge D K^{a_3}{}_{c}\wedge K^{ca_4}
\Big)\wedge e^{a_5}\wedge\cdots\wedge e^{a_D},
\end{align}
where $K^{ab}$ is the contortion 1-form and $D$ denotes the exterior covariant 
derivative. The scalar $T_G$ in components is given equivalently by the 
coordinate expression reported as Eq.~(55) in \cite{Kofinas:2014owa}. 
The fundamental identity is that the Levi-Civita Gauss-Bonnet scalar $\bar G$ 
(constructed from the Levi-Civita connection) and $T_G$ differ by a total 
derivative,  namely $\bar G=-T_G+\nabla_M(\cdots)$, generalizing the TEGR 
relation $\bar R=-T+\nabla_M(\cdots)$.

Hence, we can proceed by writing  the $5$D modified teleparallel action as
\begin{align}\label{eq7}
S=\int d^5x\,e\left[\frac{1}{4}\,f(T,T_G)+\mathcal{L}_m
\right],
\end{align}
where  $f(T,T_G)$ is an arbitrary smooth function.
By varying the action with respect to the tetrad,  we obtain the field 
equations in the form \cite{Kofinas:2014owa}
\begin{align}\label{eq10}
\mathcal{E}_{ab}=4\,\Theta_{ab},
\end{align}
where the left-hand side can be written compactly as
\begin{align}\label{eq11}
\mathcal{E}_{ab}
&=
2\Big(H_{[ac]b}+H_{[ba]c}-H_{[cb]a}\Big)_{,}{}^{c}
+2\Big(H_{[ac]b}+H_{[ba]c}-H_{[cb]a}\Big)\,C^{d}{}_{dc}
\nonumber\\
&
+\Big(2H_{[ac]d}+H_{dca}\Big)\,C_{b}{}^{cd}
+4H_{[db]c}\,C_{a}{}^{dc}
+T_{acd}\,H^{cd}{}_{b}
-h_{ab}
+\Big(f-Tf_T-T_G f_{T_G}\Big)\eta_{ab},
\end{align}
where \(f_T\equiv\partial f/\partial T\), \(f_{T_G}\equiv\partial f/\partial 
T_G\), and \(H^{abc}\), and \(h_{ab}\) are the ``gravitational momenta'' 
associated 
with \(T\) and \(T_G\). In particular, we have 
\begin{align}\label{eq12}
H^{abc}=f_T\,H^{abc}_{(T)}+f_{T_G}\,H^{abc}_{(G)}, 
\end{align}
where $h_{ab}=f_{T_G}\,h^{(G)}_{ab}$ and the scalar-torsion part  is written 
explicitly as
\begin{align}\label{eq13}
H^{abc}_{(T)}=\eta^{ac}K^{bd}{}_{d}-K^{bca},
\end{align}
while the Gauss-Bonnet teleparallel part is obtained from the TEGB invariant 
\(T_G\) 
by functional differentiation with respect to the anholonomy coefficients
(\(C^{a}{}_{bc}\)) and the fünfbein, namely  \cite{Kofinas:2014owa}
\begin{align}
H^{abc}_{(G)}:=&\frac{1}{e}\left[
\frac{\partial\big(e\,T_G\big)}{\partial C_{abc}}
-\partial_M\!\left(\frac{\partial\big(e\,T_G\big)}{\partial(\partial_M 
C_{abc})}\right)
\right],
\label{eq14}
\\
h^{(G)}_{ab}
:=&\frac{1}{e}\,e_{bM}\,\frac{\delta\big(e\,T_G\big)}{\delta e^{a}{}_{M}},
\label{eq15}
\end{align}
with $C^{a}{}_{bc}=-T^{a}{}_{bc}$ in Weitzenb\"ock gauge. 
Finally, we have defined the matter energy-momentum in mixed and tangent form as
\begin{align}\label{eq9}
\Theta_a{}^{M}:=\frac{1}{e}\frac{\delta(e\mathcal{L}_m)}{\delta e^{a}{}_{M}},
\end{align}
where $\Theta_{ab}:=\Theta_a{}^{M}e_{bM}$.

We mention  here that since in this work we focus on $D=5$, expression 
(\ref{eq6}) provides the specific expression
\begin{align}
&T_G\,e^{0}\wedge e^{1}\wedge e^{2}\wedge e^{3}\wedge e^{4}
\nonumber\\
&=
\epsilon_{abcde}
\Big[
K^{a}{}_{f}\wedge K^{fb}\wedge K^{c}{}_{g}\wedge K^{gd}
-2\,K^{ab}\wedge K^{c}{}_{f}\wedge K^{fg}\wedge K_{g}{}^{d}
+2\,K^{ab}\wedge D K^{c}{}_{f}\wedge K^{fd}
\Big]\wedge e^{e},
\label{eq16}
\end{align}
with \(K^{ab}=K^{ab}{}_{c}\,e^{c}\) the contortion 1-form and \(D\) the 
exterior covariant derivative with the teleparallel connection (in 
Weitzenb\"ock gauge \(D\to d\) on tangent scalars/forms).

Equations \eqref{eq11} together with \eqref{eq13},
\eqref{eq14}-\eqref{eq15}, and the explicit TEGB definition \eqref{eq16},
constitute the complete fünfbein equations of motion of five-dimensional 
$f(T,T_G)$ gravity.

\subsection{Thick brane ansatz and explicit invariants}
\label{subsec:brane_ansatz}

Let us now proceed to the 
 construction of braneworld configurations within $f(T,T_G)$ gravity. We 
consider the standard five-dimensional warped geometry
\begin{align}\label{eq17}
ds^2=e^{2A(y)}\eta_{\mu\nu}dx^\mu dx^\nu+dy^2,
\end{align}
with $\mu,\nu=0,1,2,3$, and the diagonal fünfbein
\begin{align}\label{eq18}
e^{a}{}_{M}=\mathrm{diag}\big(e^{A(y)},e^{A(y)},e^{A(y)},e^{A(y)},1\big).
\end{align}
This ansatz provides a natural framework for describing thick branes, where the 
warp factor $A(y)$ encodes the localization of fields along the extra dimension.

A direct computation of the torsion scalar yields
\begin{align}\label{eq19}
T=-12\,A'(y)^2,
\end{align}
while the teleparallel Gauss-Bonnet invariant, using Eq.~(\ref{eq6}) or the 
component expression of~\cite{Kofinas:2014owa}, takes the form
\begin{align}\label{eq20}T_G
=24\,A'^2\bigl(4A''+5A'^2\bigr)
=96A'^2A''+120A'^4.
\end{align}
These expressions explicitly show that both $T$ and $T_G$ are fully determined 
by the warp factor and its derivatives, allowing the function $f(T,T_G)$ and 
its derivatives to be expressed as functions of $y$.

In order to generate a thick brane, we introduce a canonical scalar field 
$\phi(y)$ with Lagrangian
\begin{align}\label{eq21}
\mathcal{L}_m
=-\frac12\,g^{MN}\partial_M\phi\,\partial_N\phi
-V(\phi).
\end{align}
The corresponding energy-momentum tensor reads
\begin{align}
T_{\mu\nu}^{(m)}=&-e^{2A}\eta_{\mu\nu}\left(\frac12\phi'^2+V\right),
\label{eq22}
\\
T_{yy}^{(m)}=&\frac12\phi'^2-V,
\label{eq23}
\end{align}
while variation with respect to $\phi$ yields the scalar-field equation
\begin{align}
\phi''+4A'\phi'=\frac{dV}{d\phi}.
\label{eq24}
\end{align}

Specializing the general $f(T,T_G)$ field equations of~\cite{Kofinas:2014owa} 
to the warped ansatz (\ref{eq17}) and fünfbein (\ref{eq18}), one obtains two 
independent gravitational equations, which can be chosen as the $(yy)$ and 
$(\mu\nu)$ components. In compact form, they read
\begin{align}
-f-24A'^2 f_T+4T_G f_G
-96e^{-4A}\!\left(e^{4A}A'^3 f_G\right)'
&=2\bigl(\phi'^2-2V\bigr),
\label{eq25_compact}
\\
f-e^{-4A}\!\left\{
e^{4A}\!\left[-6A' f_T+24A'(2A''+5A'^2)f_G\right]
\right\}'
+24e^{-4A}\!\left(e^{4A}A'^2 f_G\right)''
&=2\bigl(\phi'^2+2V\bigr).
\label{eq26_compact}
\end{align}

Using Eq.~\eqref{eq20}, the first equation can be expanded as
\begin{align}\label{eq25}
-f-24A'^2 f_T
+96A'^2(A''+A'^2)f_G
-96A'^3 f_G'
=2\bigl(\phi'^2-2V\bigr),
\end{align}
while the second becomes
\begin{align}\label{eq26}
&f+24A'^2 f_T
+6A'' f_T
+6A' f_T'
-96A'^4 f_G
\nonumber\\&-72A'^2A'' f_G
+\bigl(72A'^3+48A'A''\bigr)f_G'
+24A'^2 f_G''
=2\bigl(\phi'^2+2V\bigr).
\end{align}
Thus, Eqs.~\eqref{eq24}-\eqref{eq26} form a closed system of ordinary differential equations for the background functions $A(y)$ and $\phi(y)$ once the model $f(T,T_G)$ and the scalar potential $V(\phi)$ are specified. Since $f_T$ and $f_G$ depend on $y$ through $T(y)$ and $T_G(y)$, their derivatives can be written as
\begin{align}\label{eq27}
f_T' = f_{TT}\,T' + f_{TG}\,T_G',
\qquad
f_G' = f_{GT}\,T' + f_{GG}\,T_G',
\end{align}
and
\begin{align}\label{eq28}
f_G''
=\frac{d}{dy}\left(f_G'\right)
=f_{GTT}\,T'^2+2f_{GTG}\,T'T_G'+f_{GGG}\,T_G'^2+f_{GT}\,T''+f_{GG}\,T_G''.
\end{align}
The required derivatives of the invariants are
\begin{align}\label{eq29}
T'=-24A'A'',
\qquad
T''=-24\left[(A'')^2+A'A'''\right],
\end{align}
and
\begin{align}\label{eq30}
T_G'
&=480A'^3A''+192A'(A'')^2+96A'^2A''',
\\
T_G''
&=480A'^3A'''+1440A'^2(A'')^2+576A'A''A'''+192(A'')^3+96A'^2A''''.
\end{align}
These identities are useful for rewriting the system purely in terms of $A(y)$ and its derivatives.

It is instructive to consider the limiting cases. In particular, in the 
teleparallel equivalent of General Relativity, corresponding to $f(T,T_G)=-T$, 
one has $f_T=-1$ and $f_{T_G}=0$, and the above system reduces to the standard 
Einstein-scalar thick-brane equations 
\cite{Randall:1999vf,Randall:1999ee,Gremm:1999pj,Bazeia:2008zx,
Dzhunushaliev:2009va}:
\begin{align}
6A'^2=\phi'^2-2V, \\
3A''+6A'^2=-\phi'^2-2V, \\
\phi''+4A'\phi'=V_{,\phi}.
\end{align}

A crucial feature of the present setup is that, in five dimensions, the 
torsional Gauss-Bonnet invariant $T_G$ contributes dynamically to the field 
equations. This contrasts with the four-dimensional case, where it is purely 
topological. Consequently, the presence of the $T_G$ sector introduces 
additional derivative structures, through $f_{T_G}'$ and $f_{T_G}''$, which can 
significantly affect the warp factor, scalar dynamics, and ultimately the 
structure of thick-brane solutions. This opens the possibility for 
qualitatively new behaviors compared to both $f(T)$ and curvature-based 
Gauss-Bonnet braneworld models.

\section{Thick brane solutions and structure}\label{s3}

In this section we proceed to construct explicit thick-brane configurations 
within the five-dimensional $f(T,T_G)$ framework derived in the previous 
section, and to analyze their geometric and physical properties. As a specific 
model we consider
\begin{eqnarray}
f(T,T_G)=-T+\alpha T_G, 
\label{modelform}
\end{eqnarray}
where $\alpha$ is the coupling parameter,
since this simple, but non-trivial in five dimensions, form will prove to lead 
to interesting and rich phenomenology.
Our goal is therefore twofold. Firstly, we obtain consistent background 
solutions by adopting a suitable warped ansatz and solving the coupled system 
for the warp factor and scalar field. Secondly, we investigate how the presence 
of the $T_G$ sector, controlled by the coupling parameter $\alpha$, modifies 
key characteristics of the brane, including its thickness, internal structure, 
and energy distribution.

\subsection{Analytical framework and assumptions}

Although the system of equations (\ref{eq25})-(\ref{eq26}) constitutes a 
well-defined second-order framework, obtaining exact analytical solutions is in 
general not feasible due to their non-linear structure. In order to proceed and 
capture the essential physical features of the model, we adopt the standard 
warped ansatz \cite{Gremm:1999pj}
\begin{eqnarray}\label{20}
e^{2A(y)}=\cosh^{-2p}(\lambda y),
\end{eqnarray}
where the parameter $p$ controls the deformation of the warp factor within the 
brane core, while $\lambda$ determines the characteristic width of the brane.
Hence, within this setup,  the form (\ref{modelform}) gives
\begin{eqnarray}
f(T,T_G(y)) 
&=&12 p^2 \lambda^2 \tanh^2(y \lambda) 
\left[ 
1 - 8 p \alpha \lambda^2 \,\text{sech}^2(y \lambda) 
+ 10 p^2 \alpha \lambda^2 \tanh^2(y \lambda) 
\right].
\end{eqnarray}

Following a procedure similar to Ref.~\cite{Yang:2012hu}, one can combine 
Eqs.~(\ref{eq25})-(\ref{eq26}) and, upon inserting the ansatz (\ref{20}), 
recast the system into
\begin{eqnarray}\label{q.5}
\phi'^2(y) = \frac{3}{4} p \lambda^2 \,\mathrm{sech}^4(y \lambda)
\Big[1 + 32\alpha p^2 \lambda^2 +(1-32\alpha p^2 \lambda^2 ) \cosh^2(2y \lambda)\Big],
\end{eqnarray}
and
\begin{eqnarray}\label{q.55}
V(\phi(y)) &=& \frac{1}{4} p \lambda^2 \Bigg\{
4 p ( 54 p^2 \alpha \lambda^2-5) + \Big[3 + 4 p \Big(5 - 36 p (2 + 3 p) \alpha \lambda^2\Big)\Big]  \,\mathrm{sech}^2(y \lambda)\nonumber\\
&+&72 \alpha  p^2\lambda^2  (4 + 3 p) \,\mathrm{sech}^4(y \lambda)
\Bigg\}.
\end{eqnarray}
Equation (\ref{q.5}) determines the scalar field profile $\phi(y)$, which can 
be obtained through direct integration. When the resulting relation is 
invertible, one may express $y$ as a function of $\phi$, allowing the potential 
to be written explicitly as $V=V(\phi)$.

\subsection{Numerical solutions}

\begin{figure}[ht]
\begin{center}
\begin{tabular}{ccc}
\includegraphics[height=5cm]{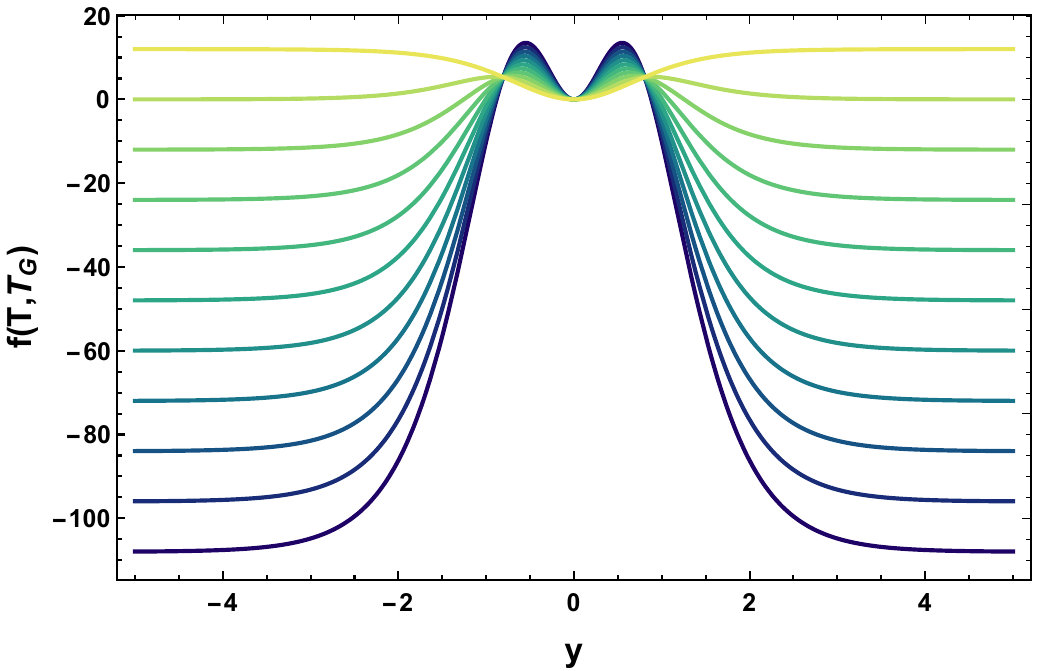}
\includegraphics[height=5cm]{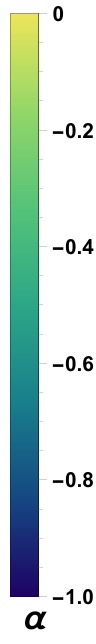}
\end{tabular}
\end{center}
\vspace{-0.5cm}
\caption{\it Behavior of the gravitational function $f(T,T_G)$ as a function of 
the extra-dimensional coordinate $y$, for $\lambda=p=1$ and representative 
values of the coupling parameter $\alpha$. The profiles are symmetric with 
respect to the brane center ($y=0$), while increasing $|\alpha|$ leads to 
pronounced deformations near the core, reflecting the nontrivial contribution 
of 
the torsional Gauss-Bonnet term.} 
\label{fig1}
\end{figure}

We elaborate the above equations numerically, and from now on all dimensional 
quantities are measured in Planck units. Fig. \ref{fig1} presents the profile 
of the gravitational function $f(T,T_G)$ as a function of the extra-dimensional 
coordinate $y$, for $\lambda=p=1$ and various values of the coupling parameter 
$\alpha$. The configuration is symmetric with respect to the brane center 
($y=0$), reflecting the underlying $\mathbb{Z}_2$ symmetry of the warped 
geometry. In the limit $\alpha \to 0$, the solution reduces to the standard 
teleparallel case, while non-zero values of $\alpha$ introduce significant 
deformations due to the torsional Gauss-Bonnet term.  
In particular, negative values of $\alpha$ enhance the amplitude of the profile 
around the brane core, indicating a stronger localization of the gravitational 
sector. This behavior reveals that the torsional Gauss-Bonnet contribution 
plays a non-trivial role in shaping the geometry of the brane, effectively 
modifying its thickness and the distribution of gravitational effects along the 
extra dimension.

\begin{figure}[ht!]
\begin{center}
\begin{tabular}{ccc}
\includegraphics[height=5cm]{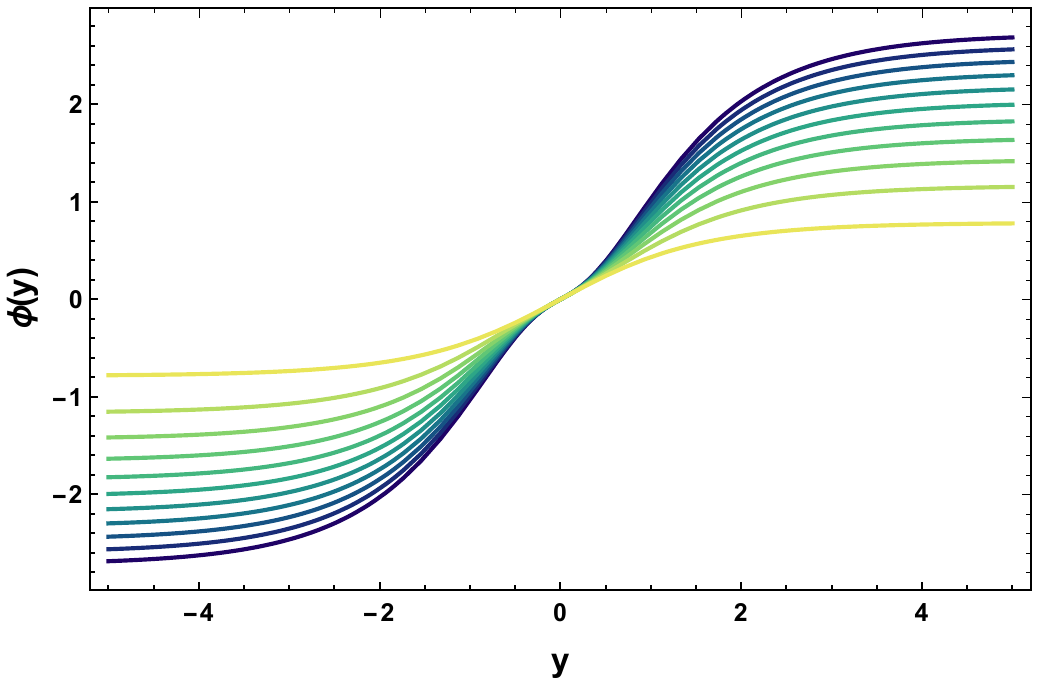} 
\includegraphics[height=5cm]{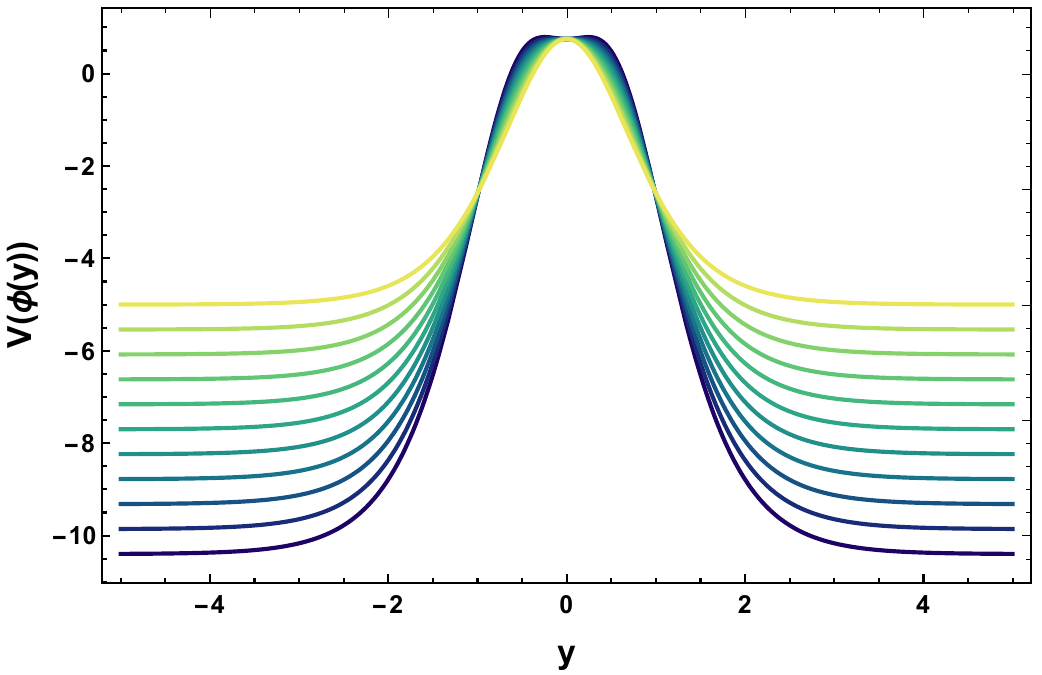}\includegraphics[height=5.1cm]{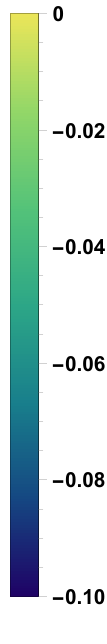}
\end{tabular}
\end{center}
\vspace{-0.5cm}
\caption{\it Profiles of the scalar field $\phi(y)$ (left panel) and the 
corresponding potential $V(\phi(y))$ (right panel), for $\lambda=p=1$ and 
representative values of the coupling parameter $\alpha$. The scalar field 
exhibits a kink-like configuration, supporting the formation of a domain-wall 
brane, while variations of $\alpha$ modify the steepness of the transition and 
the depth of the potential, indicating the impact of the torsional Gauss-Bonnet 
sector on the brane thickness and scalar dynamics.}
\label{fig2}
\end{figure}

Fig. \ref{fig2} displays the scalar field configuration $\phi(y)$ and the 
corresponding potential $V(\phi(y))$, for $\lambda=p=1$ and different values of 
$\alpha$. The scalar field exhibits the typical kink-like behavior, smoothly 
interpolating between two asymptotic constant values as $y \to \pm \infty$, 
thus realizing a domain-wall brane configuration. The presence of the torsional 
Gauss-Bonnet term significantly affects the 
scalar sector. In particular, variations of $\alpha$ modify the steepness of 
the kink profile, indicating that higher-order torsional contributions directly 
influence the internal structure and thickness of the brane. The potential 
exhibits a well-like profile centered at the origin, whose depth and curvature 
are strongly controlled by $\alpha$. For decreasing $\alpha$, the potential 
becomes deeper and narrower, suggesting an enhanced confinement mechanism for 
the scalar field around the brane core.

\subsection{Energy density and brane splitting}

We now proceed to the calculation of the  energy density and pressure. From 
(\ref{eq22}),(\ref{eq23})  we find
\begin{eqnarray}
\rho(y)&=&\frac{1}{2} p \lambda^2 \Bigg\{108 p^3 \alpha \lambda^2 \tanh^4(y 
\lambda)
+3 \,\mathrm{sech}^2(y \lambda)
- 2 p \Big[1 + 96 p \alpha \lambda^2 \,\mathrm{sech}^2(y 
\lambda)\Big]\tanh^2(y \lambda) 
\Bigg\},
\end{eqnarray}
while the pressure reads
\begin{eqnarray}
P(y)= p^2 \lambda^2 \tanh^2(y \lambda)\Bigg\{6 p \alpha \lambda^2\Big[9p-(8+9p)\mathrm{sech}^2(y \lambda)\Big] \tanh^2(y 
\lambda)-5\Bigg\}.
\end{eqnarray}

\begin{figure}[ht!]
\begin{center}
\begin{tabular}{ccc}
\includegraphics[height=5cm]{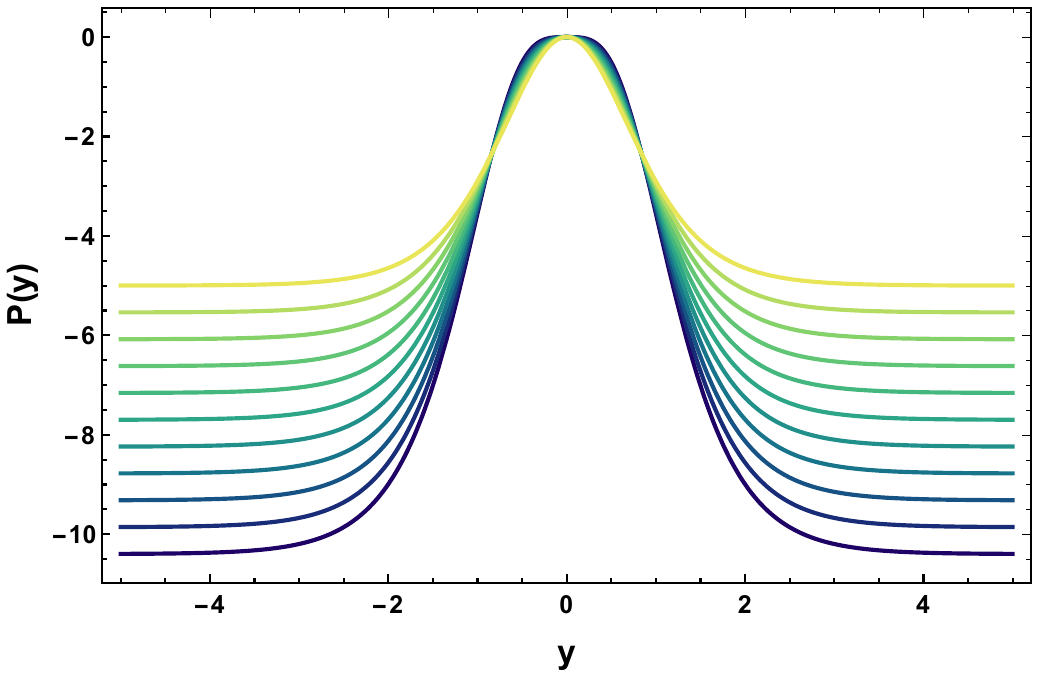}
\includegraphics[height=5cm]{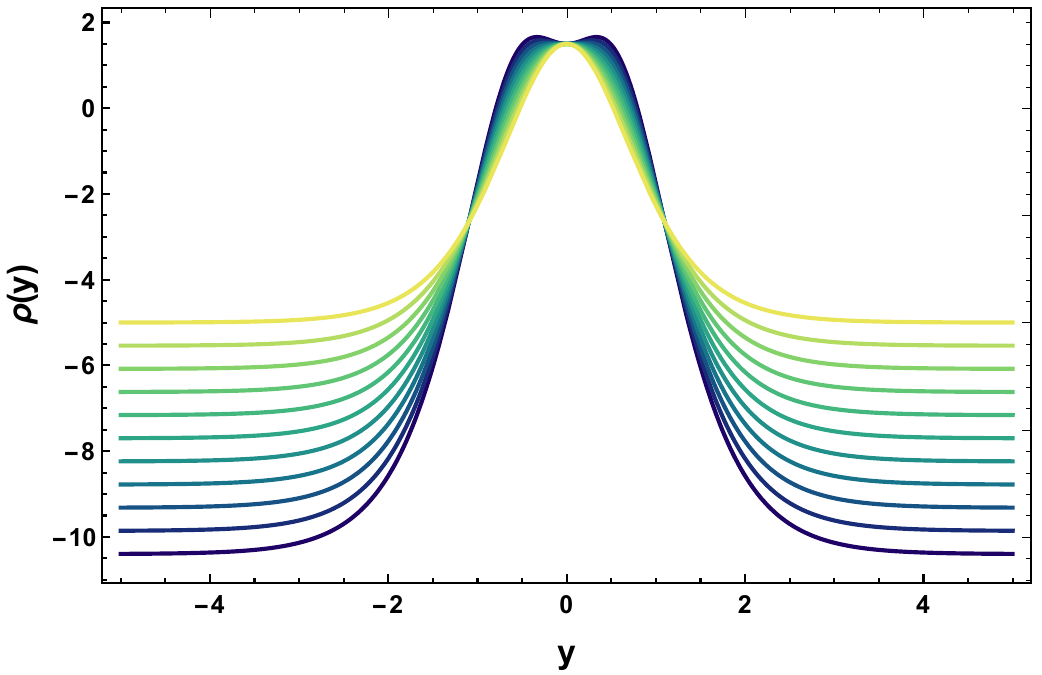}\includegraphics[height=5cm]{fig0.pdf}\\
\end{tabular}
\end{center}
\vspace{-0.5cm}
\caption{\it Profiles of the pressure $P(y)$ (left panel) and the energy 
density $\rho(y)$ (right panel), for $\lambda=p=1$ and representative values of 
the coupling parameter $\alpha$. The pressure remains negative and localized 
around the brane, ensuring stability of the configuration, while the energy 
density is peaked at the brane center. For suitable values of $\alpha$, a 
transition from a single-peak to a double-peak structure is observed, 
indicating the emergence of brane splitting and nontrivial internal structure 
induced by the torsional Gauss-Bonnet term.}
\label{fig3}
\end{figure}

Fig. \ref{fig3} shows the corresponding profiles of the energy density 
$\rho(y)$ and pressure $P(y)$, for $\lambda=p=1$ and various values of 
$\alpha$. The pressure remains negative throughout the extra dimension and 
approaches constant values asymptotically, a behavior characteristic of stable 
thick brane configurations where negative pressure counterbalances gravitational 
effects.
On the other hand, the energy density exhibits a localized structure around the 
brane core at $y=0$. Remarkably, for certain values of $\alpha$, a transition 
from a single-peak to a double-peak profile is observed, signaling the 
emergence of brane splitting and internal structure. This feature demonstrates 
that the torsional Gauss-Bonnet term induces qualitative modifications in the 
matter distribution, allowing for richer configurations compared to the 
standard single-kink scenario.

\section{Fermion localization and resonances}\label{s4}

Having established the thick-brane background solutions in the previous 
section, we now turn to the localization properties of spin-$1/2$ fermions 
propagating in this geometry. In braneworld scenarios, the ability to localize 
fermionic fields on the brane is a crucial requirement for reproducing 
effective four-dimensional physics, and it is highly sensitive to both the 
background geometry and the scalar-field configuration.

In the present $f(T,T_G)$ framework, the modified gravitational 
dynamics, arising from the torsional Gauss-Bonnet sector, affects the warp 
factor and scalar profile, and thus can have a nontrivial impact on the 
localization mechanism. Our aim is therefore to investigate how these torsional 
higher-order corrections influence the fermionic sector, including the 
existence of localized zero modes and the structure of the massive Kaluza-Klein 
spectrum. 
To achieve localization, we introduce a Yukawa-type coupling between the 
fermion field and the background scalar field, which provides the necessary 
mechanism for trapping chiral fermions on the brane.

\subsection{Dirac equation and chiral decomposition}

We consider the following five-dimensional Dirac action:
\begin{equation}\label{ad6}
 S=\int d^5x\, h \,\overline{\psi}\left[\Gamma^M D_M - G(\phi)\right]\psi,
\end{equation}
where $\Gamma^M = h_{a}\ ^M \gamma^{a}$ are the Dirac matrices in curved 
spacetime, and $\gamma^{a}$ denote the flat spacetime Dirac matrices. Moreover, 
the covariant derivative is defined as $D_M=\partial_M+\Omega_M$. 
The Yukawa coupling function $G(\phi)$ is introduced in order to ensure the 
normalizability of fermionic zero modes, a necessary condition for localization 
on the brane. Similar couplings have been widely employed in thick brane 
scenarios to control the interaction between fermions and the background scalar 
field.

Since we are working within the teleparallel framework, the spin connection 
$\Omega_M$ is expressed in terms of the contortion tensor in the Weitzenb\"ock  
gauge as
\begin{equation}
 \Omega_M=\frac{1}{4}(K_M\ ^{ab})\gamma_{a}\gamma_{b}.
\end{equation}
Additionally, the equation of motion for the fermion field is then given by
\begin{equation}\label{em11}
[\Gamma^M D_M - G(\phi)]\psi=0.
\end{equation}
For the metric (\ref{eq17}), the non-vanishing components of the spin 
connection are
\begin{eqnarray}
\Omega_\mu&=&\frac{1}{4}A^{\prime}e^{A}\gamma_{\mu}\gamma_4, \nonumber\\
\Omega_y&=&A^{\prime},
\end{eqnarray}
and thus Eq.~(\ref{em11}) takes the form
\begin{equation}\label{dir}
 e^{-A}\gamma^\mu\partial_\mu\psi + 
\gamma^4\left(\partial_y+2A^{\prime}\right)\psi + G(\phi)\psi=0.
\end{equation}
Finally, in order to simplify the analysis, we introduce the conformal 
coordinate $z$ defined through $dz = e^{-A(y)} dy$. In terms of this coordinate, 
the Dirac equation becomes
\begin{eqnarray}\label{7}
\left[\gamma^{\mu}\partial_\mu + \gamma^4(\partial_z+2\dot{A}) - 
G(\phi)\right]\psi=0,
\end{eqnarray}
where a dot denotes differentiation with respect to $z$.

We now perform a Kaluza-Klein (KK) decomposition of the fermion field:
\begin{equation}\label{kkf} 
\psi=\sum_n\left[\psi_R^{(n)}(x^\mu)\chi_{R}^{(n)}(z)+\psi_L^{(n)}(x^\mu)\chi_{L
}^{(n)}(z)\right],
\end{equation}
where $\psi_{R}^{(n)}$ and $\psi_{L}^{(n)}$ correspond to the right- and 
left-chiral components, respectively. 
We adopt a representation in which the four-dimensional spinors satisfy
\[
\gamma^\mu\partial_\mu\psi_{R,L}^{(n)} = m_n \psi_{L,R}^{(n)}, 
\quad \gamma^4 \psi_{R,L}^{(n)} = \pm \psi_{R,L}^{(n)}.
\]
Substituting the decomposition (\ref{kkf}) into Eq.~(\ref{7}), we obtain the 
coupled system
\begin{eqnarray}\label{efd}
\dot{\chi}_{R}^{(n)} + e^{A}G(\phi)\chi_{R}^{(n)} &=& 
-m_n\chi_{L}^{(n)},\nonumber\\
\dot{\chi}_{L}^{(n)} - e^{A}G(\phi)\chi_{L}^{(n)} &=& m_n\chi_{R}^{(n)}.
\end{eqnarray}
These equations can be decoupled into Schrödinger-like equations of the form
\begin{eqnarray}\label{10}
-\ddot{\chi}_{L}^{(n)}+ V_L(z)\chi_{L}^{(n)} &=& m_n^2 
\chi_{L}^{(n)},\nonumber\\
-\ddot{\chi}_{R}^{(n)}+ V_R(z)\chi_{R}^{(n)} &=& m_n^2 \chi_{R}^{(n)},
\end{eqnarray}
where the effective potentials are given by
\begin{equation}
V_{R,L}(z) = U^2(z) \pm \partial_z U(z),
\end{equation}
with
\begin{equation}
U(z) = e^A G(\phi(z)).
\end{equation}
The above structure ensures the stability of the fermionic sector, since the 
Hamiltonians can be factorized in a supersymmetric quantum mechanics form, thus 
preventing the appearance of tachyonic modes.
Finally, we consider a general Yukawa coupling of the form
\begin{eqnarray}
G(\phi)=\xi\phi^\beta,
\end{eqnarray}
which allows us to explore different localization scenarios depending on the 
choice of the parameters. In the following, we will focus on the representative 
cases $\beta=1$ and $\beta=2$.

\subsection{Zero modes}

The Schrödinger-type equations in Eq.~(\ref{10}), governing the chiral 
components of the fermion field, admit a natural interpretation within the 
framework of supersymmetric quantum mechanics. To this end, we introduce the 
first-order operators
\begin{equation}
\mathcal{A} = \partial_z + U(z), \qquad 
\mathcal{A}^{\dagger} = -\partial_z + U(z),
\end{equation}
which allow the corresponding Hamiltonians to be factorized as
\begin{equation}
H_L = \mathcal{A}^{\dagger}\mathcal{A}, \qquad 
H_R = \mathcal{A}\mathcal{A}^{\dagger}.
\end{equation}
This structure ensures that the mass spectrum is non-negative, $m^2 \geq 0$, 
and thus the fermionic sector is free from tachyonic instabilities. 
Consequently, the background configuration is stable under fermionic 
perturbations, and the associated Kaluza-Klein tower is physically well-defined.

Within this framework, the zero-mode solutions ($m=0$) can be obtained 
analytically. Their profiles are given by
\begin{equation}
\chi_{L,R}^{(0)}(z) \propto \exp\left[\pm \int U(z)\, dz \right],
\end{equation}
indicating that their localization properties are directly controlled by the 
effective function $U(z)=e^A G(\phi(z))$.
Furthermore, the localization of fermionic zero modes requires the 
normalizability condition
\begin{equation}
\int |\chi_{L,R}^{(0)}(z)|^2 dz < \infty,
\end{equation}
which determines whether a given chirality can be trapped on the brane.

\begin{figure}[!]
\begin{center}
\includegraphics[height=5cm]{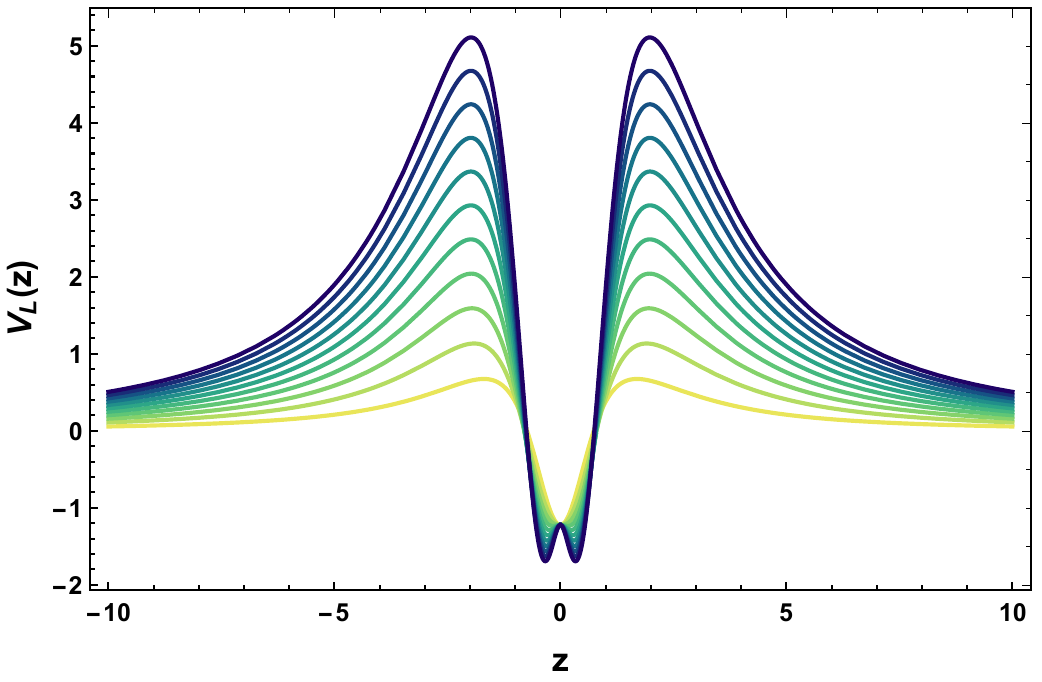}
\includegraphics[height=5cm]{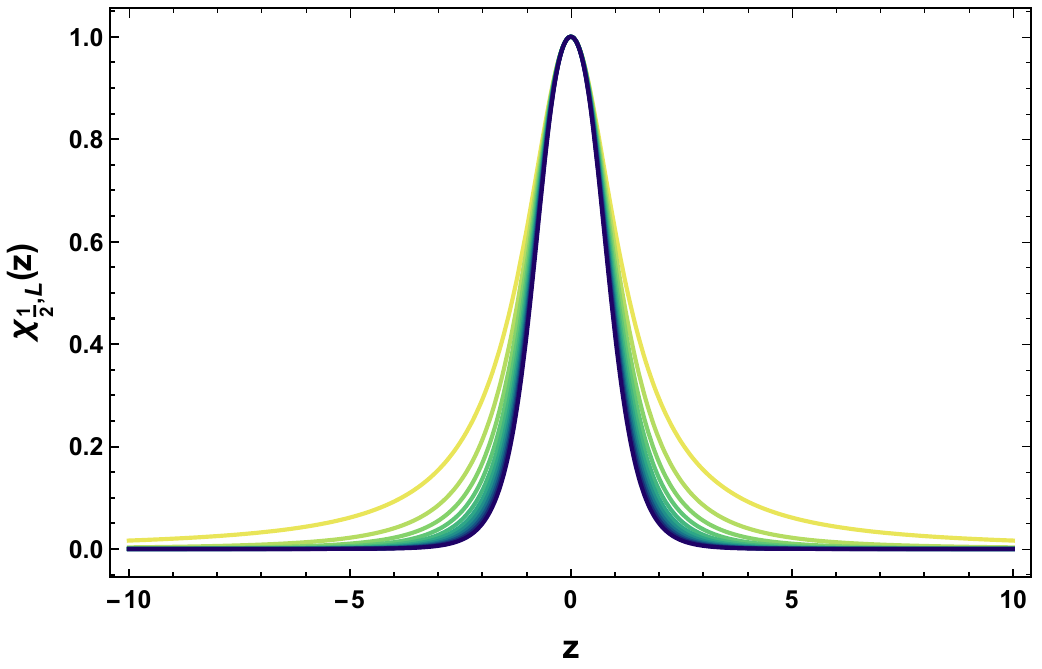} \includegraphics[height=5cm]{fig0.pdf}\\
{\it  (a) Effective potential $V_{L}(z)$ }\hspace{4cm} {\it  (b) Massless mode 
$\chi_{\frac{1}{2},L}(z)$}\\
\includegraphics[height=5cm]{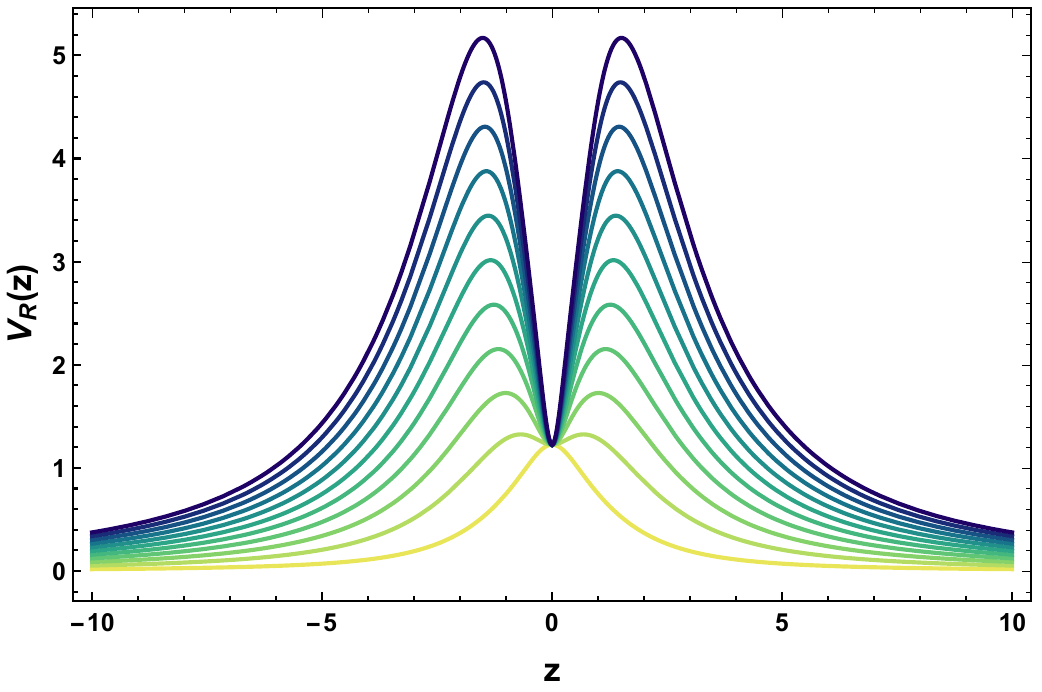}
\includegraphics[height=5cm]{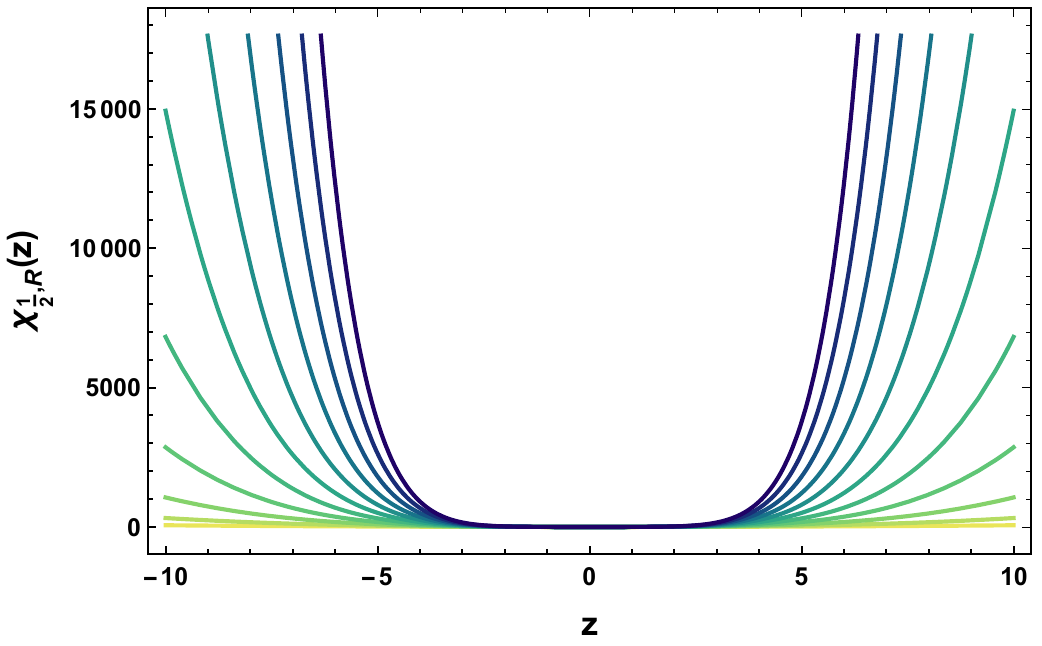}\includegraphics[height=5cm]{fig0.pdf}\\
{\it  (c)   Effective potential $V_{R}(z)$} \hspace{4cm}
{\it  (d) Massless mode 
$\chi_{\frac{1}{2},R}(z)$}
\end{center}
\vspace{-0.5cm}
\caption{\it Effective potentials and corresponding zero-mode profiles for 
spin-$1/2$ fermions, for $\xi=\lambda=p=1$ and $\beta=1$. Panels (a) and (c) 
show the left- and right-handed effective potentials $V_L(z)$ and $V_R(z)$, 
respectively, while panels (b) and (d) display the corresponding zero-mode wave 
functions $\chi_{L}^{(0)}(z)$ and $\chi_{R}^{(0)}(z)$. The left-handed 
potential exhibits a volcano-like structure that supports a normalizable zero 
mode localized on the brane, whereas the right-handed potential remains 
positive-definite and does not allow for localization. This demonstrates the 
emergence of chiral fermion localization in the present setup.}
 \label{fig4}
\end{figure}

Fig. \ref{fig4} illustrates the effective potentials $V_{L}(z)$ and 
$V_{R}(z)$, together with the corresponding zero-mode profiles 
$\chi_{L}^{(0)}(z)$ and $\chi_{R}^{(0)}(z)$, for the case $\xi=\lambda=p=1$ and 
$\beta=1$. As we observe, the left-handed potential $V_{L}(z)$ exhibits a 
characteristic volcano-like structure, with a central well surrounded by 
potential barriers. This structure supports a normalizable zero mode, whose wave 
function is sharply peaked at the brane position ($z=0$) and rapidly decays away 
from it. 
In contrast, the right-handed potential $V_{R}(z)$ remains positive-definite 
and does not develop a confining well. As a result, the corresponding 
right-handed zero mode is non-normalizable and cannot be localized on the 
brane. This leads to a chiral localization mechanism, where only one fermionic 
chirality is trapped, a feature that is essential for constructing 
phenomenologically viable braneworld models.

\begin{figure}[ht!]
\begin{center}
\begin{tabular}{ccc}
\includegraphics[height=5cm]{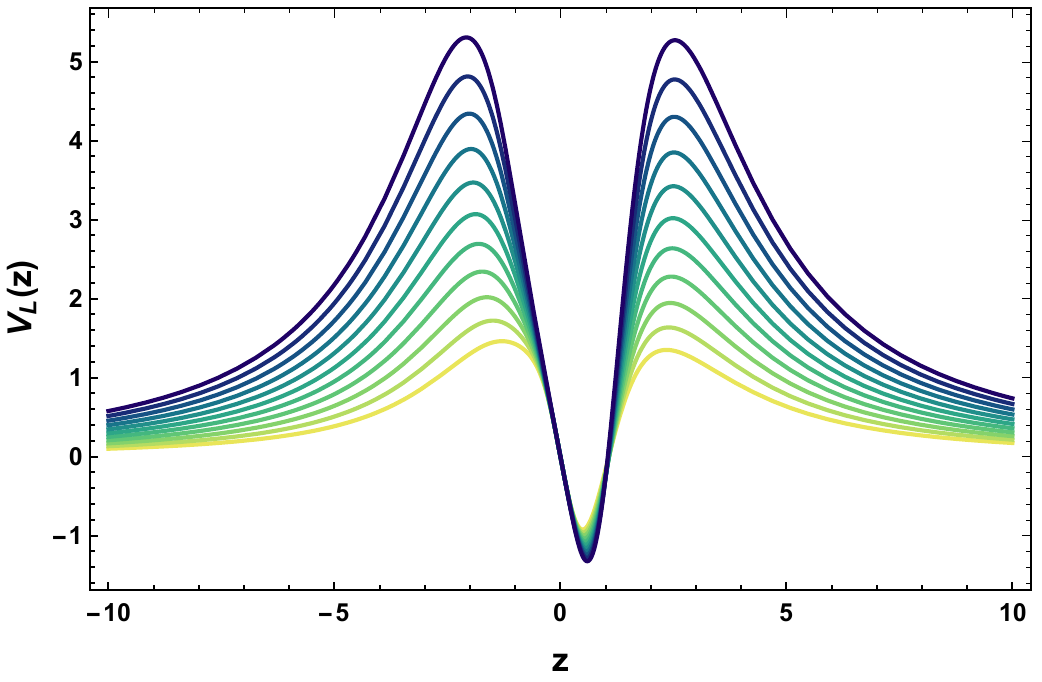}
\includegraphics[height=5cm]{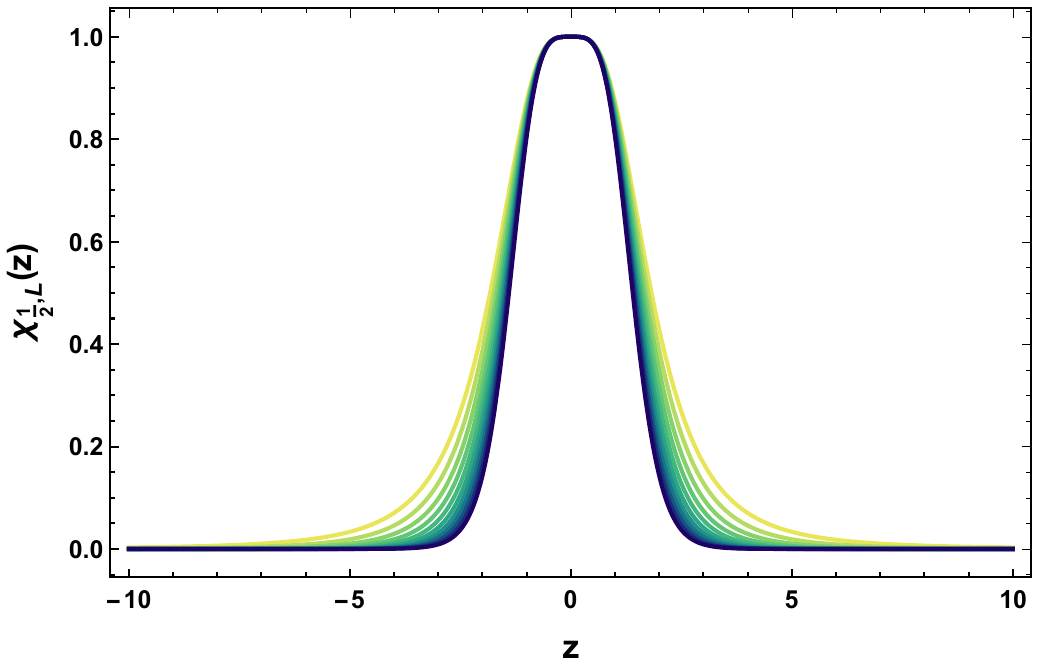} 
\includegraphics[height=5cm]{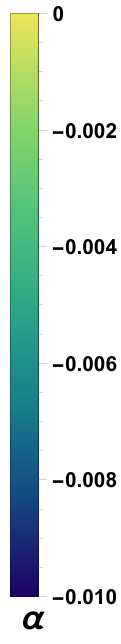}\\
{\it  (a) Effective potential $V_{L}(z)$ }\hspace{4cm} {\it   (b) Massless mode 
$\chi_{\frac{1}{2},L}(z)$ }  \\
\includegraphics[height=5cm]{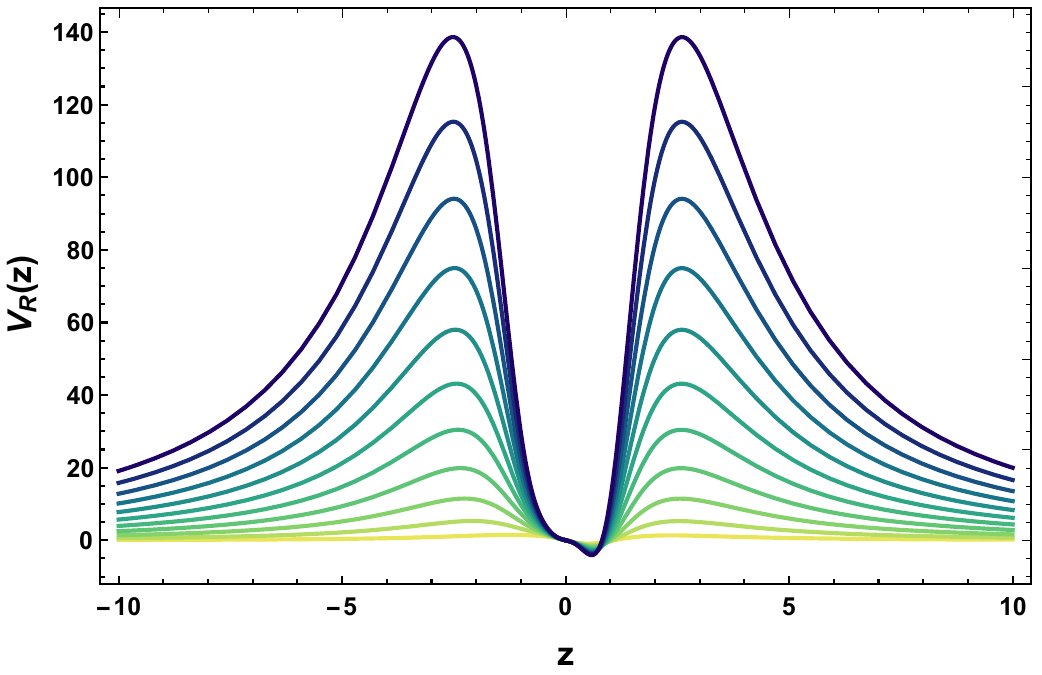}
\includegraphics[height=5cm]{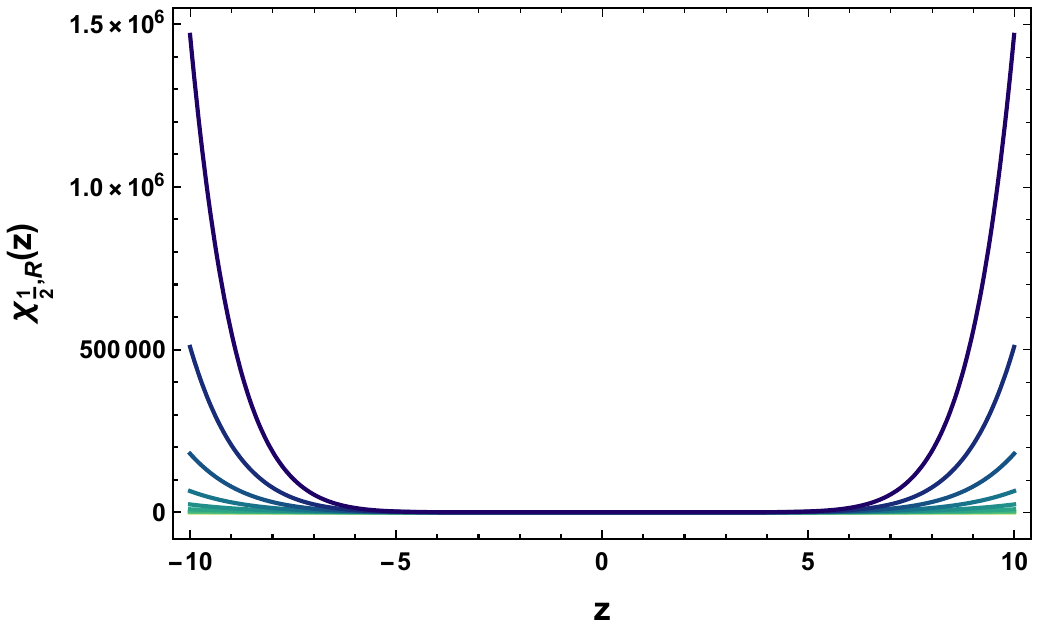}\includegraphics[height=5cm]{fig00.pdf}\\
{\it  (c)  Effective potential $V_{R}(z)$} \hspace{4cm} {\it  (d) Massless mode 
$\chi_{\frac{1}{2},R}(z)$}
\end{tabular}
\end{center}
\vspace{-0.5cm}
\caption{\it Effective potentials and corresponding zero-mode profiles for 
spin-$1/2$ fermions, for $\xi=\lambda=p=1$ and $\beta=2$. Panels (a) and (c) 
show the left- and right-handed effective potentials $V_L(z)$ and $V_R(z)$, 
respectively, while panels (b) and (d) display the corresponding zero-mode wave 
functions $\chi_{L}^{(0)}(z)$ and $\chi_{R}^{(0)}(z)$. Compared to the 
$\beta=1$ case, the left-handed potential becomes deeper and narrower, leading 
to a more strongly localized zero mode around the brane. In contrast, the 
right-handed potential remains positive-definite and does not support 
localization, further enhancing the chiral asymmetry of the fermionic sector.}
 \label{fig5}
\end{figure}

Fig. \ref{fig5} presents the corresponding results for $\beta=2$. In this 
case, the left-handed potential retains its volcano-like profile but becomes 
deeper and narrower, reflecting the stronger coupling between the fermion and 
the scalar field. Consequently, the left-handed zero mode is more tightly 
localized around the brane, exhibiting a sharper peak at $z=0$.
On the other hand, the right-handed potential increases in magnitude while 
remaining strictly positive, further suppressing the possibility of 
localization. The corresponding right-handed zero mode becomes even more 
delocalized, confirming the absence of a normalizable solution in this sector.

The above results demonstrate that the Yukawa coupling plays a crucial role in 
controlling fermion localization. In particular, increasing the power $\beta$ 
enhances the asymmetry between the chiral sectors, strengthening the 
confinement of left-handed fermions while further excluding right-handed zero 
modes. This provides a flexible mechanism for achieving chiral localization 
within the present torsional Gauss-Bonnet braneworld framework.

\subsection{Massive modes and resonances}

The massive fermionic spectrum can be obtained by solving the Schrödinger-type 
equations (\ref{10}) under appropriate boundary conditions at the origin 
\cite{Liu:2009ve,Liu:2009mga,Moreira:2021vcf}. Due to the reflection symmetry 
of 
the effective potentials $V_{L,R}(z)$ under $z \to -z$, the solutions can be 
classified according to their parity. The corresponding boundary conditions read
\begin{eqnarray}\label{ade}
\chi_{\text{even}}(0) &=& c, \quad 
\dot{\chi}_{\text{even}}(0)=0, \nonumber\\
\chi_{\text{odd}}(0) &=& 0, \quad 
\dot{\chi}_{\text{odd}}(0)=c,
\end{eqnarray}
where $c$ is a normalization constant.
These conditions allow for a consistent decomposition of the fermionic modes 
into even and odd parity sectors, which simplifies the numerical analysis of 
the spectrum.

\begin{figure}[ht]
\begin{center}
\includegraphics[height=4.5cm]{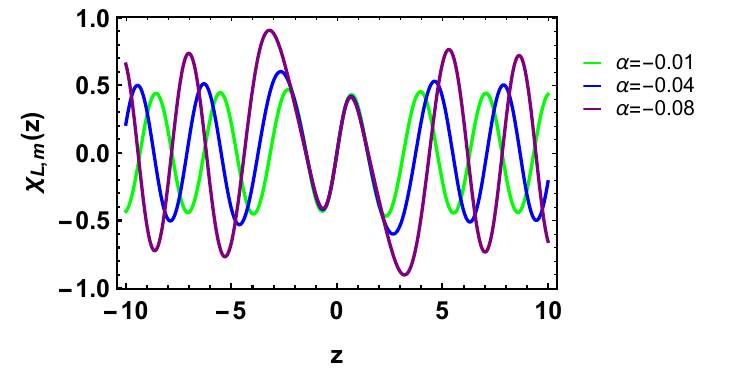}
\includegraphics[height=4.5cm]{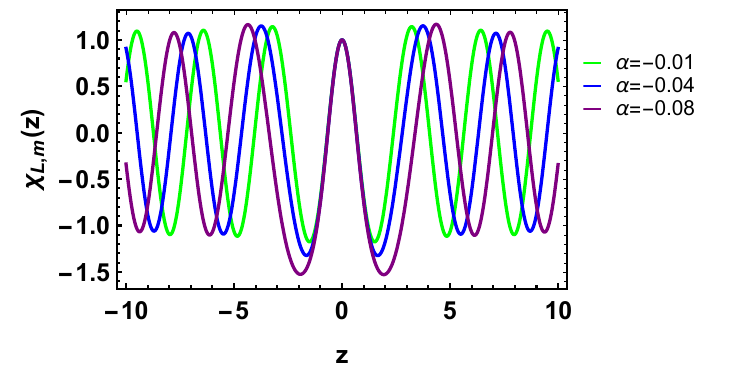}\\
\end{center}
\vspace{-0.5cm}
\caption{\it Profiles of representative massive fermionic modes for 
$\xi=\lambda=p=1$ and $\beta=1$. The left panel   corresponds to an odd mode 
with $m^2=4.7$, while the right panel   shows an even mode with 
$m^2=5.9$, for different values of the coupling parameter $\alpha$. Both 
modes exhibit oscillatory behavior along the extra dimension, characteristic of 
non-localized Kaluza-Klein states. The amplitude and phase of the oscillations 
are significantly affected by $\alpha$, with larger $|\alpha|$ enhancing the 
modulation near the brane, indicating the influence of the torsional 
Gauss-Bonnet term on the fermionic spectrum.}
\label{fig6}
\end{figure}

Fig. \ref{fig6} displays representative massive fermionic modes for 
$\beta=1$, considering both parity sectors. The left panel corresponds to an 
odd mode with $m^2=4.7$, while the right panel shows an even mode with 
$m^2=5.9$, for various values of the coupling parameter $\alpha$. As 
expected, both modes exhibit oscillatory behavior.
Nevertheless, the presence of the torsional Gauss-Bonnet coupling $\alpha$ 
significantly affects the structure of these modes. In particular, increasing 
the magnitude of negative $\alpha$ enhances the amplitude of oscillations near 
the brane core, indicating a stronger interaction between the fermionic modes 
and the modified gravitational background. This effect reflects the influence 
of the $T_G$ sector on the effective potentials, which alters the propagation 
of massive fermions in the bulk.

Additionally, as we   observe, the parity properties are clearly preserved: odd 
modes vanish at the origin, while even modes remain finite and symmetric, in 
agreement with the imposed boundary conditions. Importantly, the deformation of 
the wave profiles induced by $\alpha$ suggests that the torsional Gauss-Bonnet 
term modifies the spectral distribution of massive states, potentially affecting 
the emergence of quasi-localized configurations.

\begin{figure}[ht!]
\begin{center}
 \includegraphics[height=4.5cm]{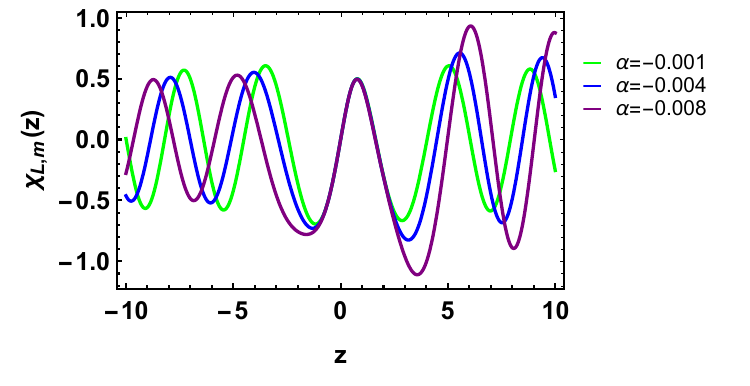}
\includegraphics[height=4.5cm]{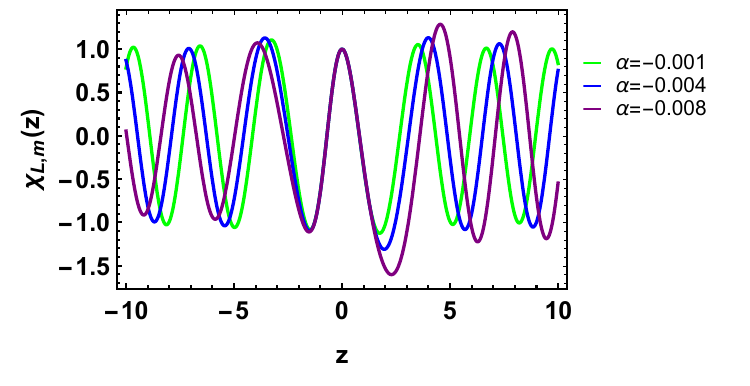}\\
 \end{center}
\vspace{-0.5cm}
\caption{\it Profiles of representative massive fermionic modes for 
$\xi=\lambda=p=1$ and $\beta=2$. The left panel   corresponds to an odd mode 
with $m^2=4.7$, while the right panel   shows an even mode with 
$m^2=5.9$, for different values of the coupling parameter $\alpha$. As in the 
$\beta=1$ case, the modes exhibit oscillatory behavior characteristic of 
non-localized Kaluza-Klein states. However, the increase in $\beta$ leads to 
enhanced oscillation amplitudes near the brane, reflecting a stronger coupling 
between the fermion and the scalar background. This results in a higher 
sensitivity of the massive spectrum to the torsional Gauss-Bonnet contribution.}
\label{fig7}
\end{figure}

Fig. \ref{fig7} presents the corresponding results for $\beta=2$. As in the 
previous case, the modes display oscillatory profiles, but with noticeably 
enhanced amplitudes. This behavior arises from the stronger Yukawa coupling, 
which intensifies the interaction between the fermionic field and the scalar 
background.
Moreover, the impact of the torsional Gauss-Bonnet term becomes more 
pronounced. Negative values of $\alpha$ lead to stronger modulations near the 
brane, indicating that higher-order torsional corrections significantly affect 
the effective potential governing fermionic dynamics. The preservation of 
parity properties remains evident, with odd modes vanishing at the origin and 
even modes exhibiting symmetric profiles. Hence, this behavior demonstrates 
that increasing the Yukawa coupling power $\beta$ 
enhances the sensitivity of the massive spectrum to the underlying geometry, 
reinforcing the role of the torsional Gauss-Bonnet contribution in shaping the 
fermionic sector.

Beyond the continuum of massive modes, it is important to examine the possible 
existence of resonant states. These correspond to quasi-localized massive modes 
that, although not strictly normalizable, exhibit a significant concentration 
of probability density near the brane. 
To identify such states, we introduce the relative probability function 
\cite{Tan:2020sys,Liu:2009mga}
\begin{eqnarray}\label{0555}
P(m)=\frac{\int_{-z_b}^{z_b} |\chi(z)|^2 \, dz}
{\int_{-z_{\text{max}}}^{z_{\text{max}}} |\chi(z)|^2 \, dz},
\end{eqnarray}
which measures the fraction of the total probability density localized within a 
finite region around the brane. The parameter $z_{\text{max}}$ defines the 
numerical cutoff, while smaller values of $z_b$ enhance the resolution of 
resonant peaks.

\begin{figure}[ht!]
\begin{center} 
\includegraphics[height=5cm]{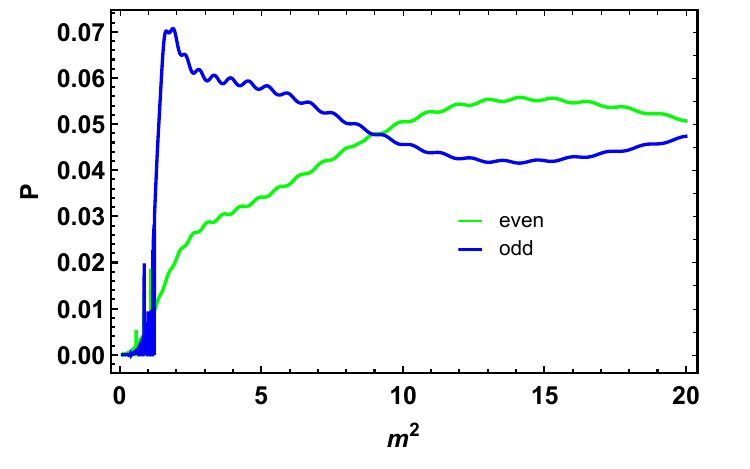} 
\includegraphics[height=5cm]{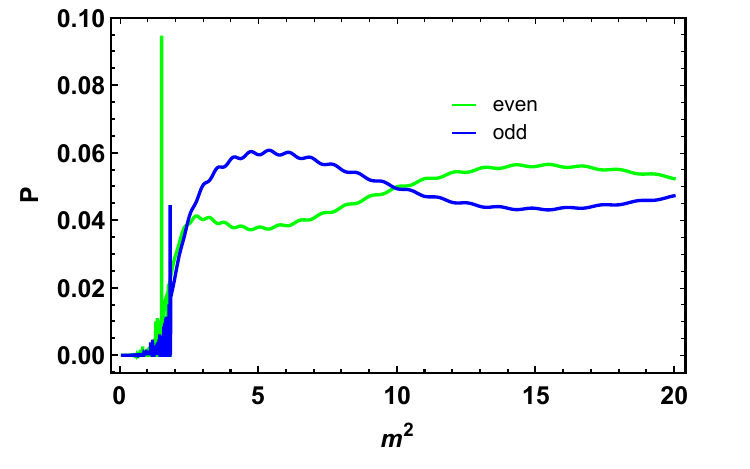}\\
\end{center}
\vspace{-0.5cm}
\caption{\it Relative probability $P(m)$ defined in Eq.~(\ref{0555}) as a 
function of the squared mass $m^2$, for $\xi=\lambda=p=1$ and $\alpha=-0.01$. 
The left panel  corresponds to $\beta=1$, while the right panel  shows 
$\beta=2$. Pronounced peaks in $P(m)$ indicate the presence of resonant states, 
corresponding to quasi-localized massive fermions near the brane. Increasing 
$\beta$ leads to a redistribution of the resonance structure, with broader and 
less sharply defined peaks, reflecting the impact of the Yukawa coupling on the 
fermionic spectrum.}
\label{fig13}
\end{figure}

Fig. \ref{fig13} shows the behavior of $P(m)$ as a function of $m^2$, for 
both even and odd modes, considering $\beta=1$ (left panel) and $\beta=2$ 
(right panel), with fixed parameters $\xi=\lambda=p=1$ and $\alpha=-0.01$. The 
presence of pronounced peaks at low masses signals the existence of resonant 
states, corresponding to quasi-localized fermions with enhanced probability 
density near the brane. As shown, for $\beta=1$, the odd sector exhibits 
a sharp and dominant resonance peak, 
while the even modes display a smoother behavior, indicating weaker 
localization. On the other hand, for $\beta=2$, the resonance structure becomes 
more 
distributed, with broader peaks spanning a wider mass range. This reflects a 
redistribution of the resonant spectrum due to the stronger Yukawa coupling.
At higher masses, the relative probability tends to stabilize, indicating the 
transition to fully delocalized Kaluza-Klein modes propagating in the bulk.

In summary, these results demonstrate that both the Yukawa coupling parameter 
$\beta$ and the torsional Gauss-Bonnet contribution (through $\alpha$) play a 
crucial role in shaping the structure of the massive fermionic spectrum. In 
particular, the $T_G$ sector provides an additional mechanism for controlling 
the emergence and distribution of fermionic resonances in teleparallel thick 
brane scenarios.

\section{Physical implications of the torsional Gauss-Bonnet term}\label{s5}

In the previous sections we have constructed thick-brane solutions and analyzed 
the localization properties of fermionic fields within the five-dimensional 
$f(T,T_G)$ framework. We now proceed to discuss the physical implications of 
the torsional Gauss-Bonnet contribution, focusing on how it modifies the 
geometry, matter distribution, and fermionic spectrum.

A key feature of the present setup is that, in five dimensions, the torsional 
Gauss-Bonnet term $T_G$ contributes nontrivially to the gravitational dynamics. 
This distinguishes the model from both four-dimensional teleparallel theories, 
where $T_G$ reduces to a boundary term, and from standard $f(T)$ gravity, where 
such higher-order torsional effects are absent. As a result, the parameter 
$\alpha$, controlling the strength of the $T_G$ sector, introduces genuinely 
new degrees of freedom in the structure of the brane.
At the level of the background geometry, the presence of $T_G$ leads to 
significant modifications of the warp factor and the scalar-field 
configuration. In particular, we have shown that varying $\alpha$ can induce a 
transition from a single-peak to a double-peak energy density profile, 
signaling the emergence of brane splitting and internal structure. This 
behavior indicates that higher-order torsional corrections can effectively 
generate richer brane configurations, beyond the standard domain-wall solutions 
typically encountered in $f(T)$ or General Relativity-based models.

The influence of the torsional Gauss-Bonnet term extends naturally to the 
matter sector. Since the fermionic localization mechanism depends sensitively 
on the background geometry through the effective function $U(z)=e^A 
G(\phi(z))$, the modifications induced by $T_G$ translate directly into changes 
in the effective potentials governing the chiral modes. As a result, the 
localization properties of fermions are not only controlled by the Yukawa 
coupling, but also by the underlying torsional structure of spacetime. 
In particular, the presence of $T_G$ affects both the shape and depth of the 
effective potentials, leading to enhanced localization of zero modes and 
significant deformations of the massive spectrum. This is clearly reflected in 
the behavior of the Kaluza-Klein modes, where the coupling $\alpha$ modifies 
the amplitude and structure of the wave functions, especially in the vicinity 
of the brane.

Moreover, the torsional Gauss-Bonnet sector plays a crucial role in the 
formation of resonant states. The modification of the effective potentials 
leads to changes in the distribution and intensity of resonance peaks, as 
evidenced by the behavior of the relative probability function $P(m)$. In 
particular, stronger torsional corrections can enhance or redistribute 
quasi-localized states, providing an additional mechanism to control the 
fermionic spectrum.

It is important to emphasize that these effects arise already at the level of 
the minimal extension $f(T,T_G)=-T+\alpha T_G$. This demonstrates that even the 
minimal torsional Gauss-Bonnet contribution encapsulates the essential 
higher-order corrections responsible for the observed qualitative features.
Additionally,  we note that, although curvature-based Gauss-Bonnet gravity also 
leads to modifications in higher-dimensional braneworlds, the teleparallel 
formulation considered here is conceptually distinct, since the corrections 
originate from torsion rather than curvature. This difference is reflected in 
the structure of the field equations and, consequently, in the resulting 
physical effects. Therefore, the present analysis reveals the importance of 
exploring torsion-based extensions of gravity as a complementary approach to 
curvature-based modifications.

\section{Conclusions}\label{s6}

In this work, we have constructed and analyzed thick-brane configurations in 
the context of five-dimensional $f(T,T_G)$ modified teleparallel gravity, 
focusing on the minimal extension $f(T,T_G)=-T+\alpha T_G$. A central 
motivation of this study is that, in five dimensions, the torsional 
Gauss-Bonnet invariant $T_G$ contributes nontrivially to the gravitational 
dynamics, in contrast to the four-dimensional case where it reduces to a 
boundary term. This allows the $T_G$ sector to introduce genuinely new effects 
in the structure of braneworld solutions.

Within this framework, we derived the full system of field equations for a 
warped geometry supported by a scalar field and obtained explicit thick-brane 
solutions using a standard domain-wall ansatz. We showed that the torsional 
Gauss-Bonnet contribution, controlled by the coupling parameter $\alpha$, leads 
to significant modifications of the background geometry and matter 
distribution. In particular, the warp factor, scalar profile, and effective 
potential are all deformed in a nontrivial way, resulting in a direct control 
of the brane thickness and internal structure.

A key result of our analysis is that the inclusion of the $T_G$ sector can 
induce qualitative changes in the energy density profile, including the 
transition from a single-peak to a double-peak configuration. This signals the 
emergence of brane splitting and reveals the ability of torsional higher-order 
corrections to generate richer internal structures compared to standard $f(T)$ 
or curvature-based Gauss-Bonnet braneworld models. Importantly, these effects 
arise already at the level of the linear $T_G$ contribution, indicating that 
even minimal extensions of teleparallel gravity can lead to nontrivial 
phenomenology in higher dimensions.

We then investigated the localization of spin-$1/2$ fermions in this background 
through a Yukawa-type coupling with the scalar field. The resulting 
Schrödinger-like equations exhibit a supersymmetric structure, ensuring the 
absence of tachyonic modes and thus the stability of the fermionic sector. We 
found that the system naturally realizes chiral localization, with only the 
left-handed zero mode being normalizable and confined on the brane, while the 
right-handed mode remains delocalized.

Concerning the massive sector, we showed that the Kaluza-Klein fermionic modes 
are strongly influenced by both the Yukawa coupling parameter $\beta$ and the 
torsional Gauss-Bonnet parameter $\alpha$. In particular, the $T_G$ 
contribution modifies the effective potentials and deforms the wave functions, 
especially near the brane core, thereby altering the spectral distribution of 
massive modes. Furthermore, through the analysis of the relative probability 
function, we identified the presence of fermionic resonances, corresponding to 
quasi-localized massive states. The structure, intensity, and distribution of 
these resonances depend sensitively on both $\beta$ and $\alpha$, demonstrating 
that the torsional Gauss-Bonnet sector provides an additional and independent 
mechanism for controlling the fermionic spectrum.

In summary, our results reveal that $f(T,T_G)$ gravity offers a 
qualitatively richer framework for thick-brane constructions, where 
torsion-based higher-order corrections can simultaneously affect the geometry, 
matter distribution, and localization properties of fields. The teleparallel 
Gauss-Bonnet term, in particular, emerges as a key ingredient for generating 
new braneworld features that are not present in purely torsional or 
curvature-based models, emphasizing the importance of torsion as a fundamental 
geometrical ingredient in higher-dimensional gravity.

Several interesting directions arise from the present analysis. 
One natural extension is the study of more general functional forms of 
$f(T,T_G)$, where nonlinear contributions could further enrich the structure of 
brane solutions and their stability properties. In addition, it would be 
important to investigate the localization of other fields, such as gauge and 
tensor modes, in order to assess the full phenomenological viability of the 
model. Finally, exploring cosmological realizations of $f(T,T_G)$ braneworlds, 
as well as potential observational signatures, such as corrections to Newton's 
law or imprints in Kaluza-Klein spectra, could provide a direct link between 
torsion-based modified gravity and testable predictions. These investigations 
lie beyond the scope of the present work and are left for future projects.

\subsection*{Acknowledgments} 
SHD would like to thank the 
partial support of project 20240220-SIP-IPN, Mexico, for starting this work on 
the research stay in China. FMB would like to   express 
gratitude to the Conselho Nacional de Desenvolvimento Cient\'{i}fico e 
Tecnol\'{o}gico CNPq for grant No. 151845/2025-5.   
ENS acknowledges the contribution of the LISA CosWG, and of   COST 
Actions  CA18108  ``Quantum Gravity Phenomenology in the multi-messenger 
approach''  and  CA21136 ``Addressing observational tensions in cosmology with 
systematics and fundamental physics (CosmoVerse)''.







\begin{thebibliography}{99}


\bibitem{Kaluza:1921tu}
T.~Kaluza,
Sitzungsber. Preuss. Akad. Wiss. Berlin (Math. Phys. ) \textbf{1921}, 966-972 
(1921)
[arXiv:1803.08616 [physics.hist-ph]].

\bibitem{Klein:1926tv}
O.~Klein,
Z. Phys. \textbf{37} (1926), 895-906.


\bibitem{Arkani-Hamed:1998jmv}
N.~Arkani-Hamed, S.~Dimopoulos and G.~R.~Dvali,
Phys. Lett. B \textbf{429}, 263-272 (1998)
[arXiv:hep-ph/9803315 [hep-ph]].


\bibitem{Antoniadis:1998ig}
I.~Antoniadis, N.~Arkani-Hamed, S.~Dimopoulos and G.~R.~Dvali,
Phys. Lett. B \textbf{436}, 257-263 (1998)
[arXiv:hep-ph/9804398 [hep-ph]].








\bibitem{Sahni:2002dx}
V.~Sahni and Y.~Shtanov,
JCAP \textbf{11}, 014 (2003)
[arXiv:astro-ph/0202346 [astro-ph]].


 

\bibitem{Lidsey:2002zw}
J.~E.~Lidsey, S.~Nojiri and S.~D.~Odintsov,
JHEP \textbf{06}, 026 (2002)
[arXiv:hep-th/0202198 [hep-th]].

 
\bibitem{Shtanov:2002ek}
Y.~Shtanov and V.~Sahni,
Class. Quant. Grav. \textbf{19}, L101-L107 (2002)
[arXiv:gr-qc/0204040 [gr-qc]].



\bibitem{Bento:2002np}
M.~C.~Bento, O.~Bertolami and A.~A.~Sen,
Phys. Rev. D \textbf{67}, 063511 (2003)
[arXiv:hep-th/0208124 [hep-th]].




\bibitem{Kudoh:2003xz}
H.~Kudoh, T.~Tanaka and T.~Nakamura,
Phys. Rev. D \textbf{68}, 024035 (2003)
[arXiv:gr-qc/0301089 [gr-qc]].


\bibitem{Eiroa:2004gh}
E.~F.~Eiroa,
Phys. Rev. D \textbf{71}, 083010 (2005)
[arXiv:gr-qc/0410128 [gr-qc]].

\bibitem{Calcagni:2004bh}
G.~Calcagni,
Phys. Rev. D \textbf{69}, 103508 (2004)
[arXiv:hep-ph/0402126 [hep-ph]].



\bibitem{Tretyakov:2005en}
P.~Tretyakov, A.~Toporensky, Y.~Shtanov and V.~Sahni,
Class. Quant. Grav. \textbf{23}, 3259-3274 (2006)
[arXiv:gr-qc/0510104 [gr-qc]].




\bibitem{Majumdar:2005ba}
A.~S.~Majumdar and N.~Mukherjee,
Int. J. Mod. Phys. D \textbf{14}, 1095 (2005)
[arXiv:astro-ph/0503473 [astro-ph]].



\bibitem{Alam:2005pb}
U.~Alam and V.~Sahni,
Phys. Rev. D \textbf{73}, 084024 (2006)
[arXiv:astro-ph/0511473 [astro-ph]].

\bibitem{Lazkoz:2006gp}
R.~Lazkoz, R.~Maartens and E.~Majerotto,
Phys. Rev. D \textbf{74}, 083510 (2006)
[arXiv:astro-ph/0605701 [astro-ph]].

\bibitem{Saridakis:2007wx}
E.~N.~Saridakis,
Phys. Lett. B \textbf{661}, 335-341 (2008)
[arXiv:0712.3806 [gr-qc]].

\bibitem{Lobo:2007qi}
F.~S.~N.~Lobo,
Phys. Rev. D \textbf{75}, 064027 (2007)
[arXiv:gr-qc/0701133 [gr-qc]].


\bibitem{Koyama:2007ih}
K.~Koyama and F.~P.~Silva,
Phys. Rev. D \textbf{75}, 084040 (2007)
[arXiv:hep-th/0702169 [hep-th]].

\bibitem{Saridakis:2007cf}
E.~N.~Saridakis,
Nucl. Phys. B \textbf{808}, 224-236 (2009)
[arXiv:0710.5269 [hep-th]].

\bibitem{Schee:2008fc}
J.~Schee and Z.~Stuchlik,
Gen. Rel. Grav. \textbf{41}, 1795-1818 (2009)
[arXiv:0812.3017 [astro-ph]].


 
 


\bibitem{Setare:2008mb}
M.~R.~Setare and E.~N.~Saridakis,
JCAP \textbf{03}, 002 (2009)
[arXiv:0811.4253 [hep-th]].

\bibitem{Schmidt:2009sg}
F.~Schmidt,
Phys. Rev. D \textbf{80}, 043001 (2009)
[arXiv:0905.0858 [astro-ph.CO]].


\bibitem{Abdujabbarov:2009az}
A.~Abdujabbarov and B.~Ahmedov,
Phys. Rev. D \textbf{81}, 044022 (2010)
[arXiv:0905.2730 [gr-qc]].


 

\bibitem{Schmidt:2009sv}
F.~Schmidt,
Phys. Rev. D \textbf{80}, 123003 (2009)
[arXiv:0910.0235 [astro-ph.CO]].

\bibitem{Lombriser:2009xg}
L.~Lombriser, W.~Hu, W.~Fang and U.~Seljak,
Phys. Rev. D \textbf{80}, 063536 (2009)
[arXiv:0905.1112 [astro-ph.CO]].


 

\bibitem{Schmidt:2009yj}
F.~Schmidt, W.~Hu and M.~Lima,
Phys. Rev. D \textbf{81}, 063005 (2010)
[arXiv:0911.5178 [astro-ph.CO]].

\bibitem{Abdujabbarov:2017pfw}
A.~Abdujabbarov, B.~Ahmedov, N.~Dadhich and F.~Atamurotov,
Phys. Rev. D \textbf{96}, no.8, 084017 (2017)































        
\bibitem{Randall:1999vf}
L.~Randall and R.~Sundrum,
Phys. Rev. Lett. \textbf{83}, 4690-4693 (1999)
[arXiv:hep-th/9906064 [hep-th]].

 
     
\bibitem{Randall:1999ee}
L.~Randall and R.~Sundrum,
Phys. Rev. Lett. \textbf{83}, 3370-3373 (1999)
[arXiv:hep-ph/9905221 [hep-ph]].



\bibitem{Bazeia:2008zx}
D.~Bazeia, A.~R.~Gomes, L.~Losano and R.~Menezes,
Phys. Lett. B \textbf{671}, 402-410 (2009)
[arXiv:0808.1815 [hep-th]].


 

\bibitem{Dzhunushaliev:2009va}
V.~Dzhunushaliev, V.~Folomeev and M.~Minamitsuji,
Rept. Prog. Phys. \textbf{73}, 066901 (2010)
[arXiv:0904.1775 [gr-qc]].



 

\bibitem{Gremm:1999pj}
M.~Gremm,
Phys. Lett. B \textbf{478}, 434-438 (2000)
[arXiv:hep-th/9912060 [hep-th]].


 

\bibitem{Charmousis:2001hg}
C.~Charmousis, R.~Emparan and R.~Gregory,
JHEP \textbf{05}, 026 (2001)
[arXiv:hep-th/0101198 [hep-th]].



 
\bibitem{Arias:2002ew}
O.~Arias, R.~Cardenas and I.~Quiros,
Nucl. Phys. B \textbf{643}, 187-200 (2002)
[arXiv:hep-th/0202130 [hep-th]].


 

\bibitem{Zhong:2022wlw}
Y.~Zhong, K.~Yang and Y.~X.~Liu,
JHEP \textbf{09}, 128 (2022)
[arXiv:2206.15145 [gr-qc]].


 


\bibitem{Peyravi:2022ubf}
M.~Peyravi, S.~Nazifkar, F.~S.~N.~Lobo and K.~Javidan,
Eur. Phys. J. C \textbf{83}, no.9, 832 (2023)
[arXiv:2210.17387 [gr-qc]].



 
\bibitem{Gordin:2023nsv}
J.~E.~B.~Gordin, K.~MacDevette and J.~Bruton,
JHEP \textbf{05}, 061 (2024)
[arXiv:2311.14436 [hep-th]].


 

\bibitem{Tan:2020sys}
Q.~Tan, W.~D.~Guo, Y.~P.~Zhang and Y.~X.~Liu,
Eur. Phys. J. C \textbf{81}, no.4, 373 (2021)
[arXiv:2008.08440 [gr-qc]].


 


\bibitem{Azizi:2025kwl}
T.~Azizi and M.~Alimoradi,
Class. Quant. Grav. \textbf{43}, no.1, 015006 (2026)
[arXiv:2512.22405 [gr-qc]].


 


\bibitem{Deng:2025hfn}
W.~Deng, S.~Long, Q.~Tan, Z.~C.~Chen and J.~Jing,
thick 
JHEP \textbf{01}, 066 (2026)
[arXiv:2508.20937 [gr-qc]].

 

\bibitem{Liu:2009ve}
Y.~X.~Liu, J.~Yang, Z.~H.~Zhao, C.~E.~Fu and Y.~S.~Duan,
Phys. Rev. D \textbf{80}, 065019 (2009)
[arXiv:0904.1785 [hep-th]].

 

\bibitem{Liu:2009mga}
Y.~X.~Liu, H.~T.~Li, Z.~H.~Zhao, J.~X.~Li and J.~R.~Ren,
JHEP \textbf{10}, 091 (2009)
[arXiv:0909.2312 [hep-th]].


  
\bibitem{Aldrovandi:2013wha}
R.~Aldrovandi and J.~G.~Pereira,
{\it{Teleparallel Gravity: An Introduction}},
Springer, 2013.


\bibitem{Cai:2015emx}
Y.~F.~Cai, S.~Capozziello, M.~De Laurentis and E.~N.~Saridakis,
Rept. Prog. Phys. \textbf{79}, no.10, 106901 (2016)
[arXiv:1511.07586 [gr-qc]].




\bibitem{Chen:2010va}
S.~H.~Chen, J.~B.~Dent, S.~Dutta and E.~N.~Saridakis,
Phys. Rev. D \textbf{83}, 023508 (2011)
[arXiv:1008.1250 [astro-ph.CO]].

\bibitem{Bengochea:2010sg}
G.~R.~Bengochea,
Phys. Lett. B \textbf{695}, 405-411 (2011)
[arXiv:1008.3188 [astro-ph.CO]].

 


\bibitem{Karami:2010bys}
K.~Karami and A.~Abdolmaleki,
Res. Astron. Astrophys. \textbf{13}, 757-771 (2013)
[arXiv:1009.2459 [gr-qc]].
  

  
\bibitem{HamaniDaouda:2011iy}
M.~Hamani Daouda, M.~E.~Rodrigues and M.~J.~S.~Houndjo,
Eur. Phys. J. C \textbf{71}, 1817 (2011)
[arXiv:1108.2920 [astro-ph.CO]].



 

\bibitem{Meng:2011ne}
X.~h.~Meng and Y.~b.~Wang,
Eur. Phys. J. C \textbf{71}, 1755 (2011)
[arXiv:1107.0629 [astro-ph.CO]].

\bibitem{Karami:2012fu}
K.~Karami and A.~Abdolmaleki,
JCAP \textbf{04}, 007 (2012)
[arXiv:1201.2511 [gr-qc]].

 
 
\bibitem{Tamanini:2012hg}
N.~Tamanini and C.~G.~Boehmer,
Phys. Rev. D \textbf{86}, 044009 (2012)
[arXiv:1204.4593 [gr-qc]].
  

  


\bibitem{Cardone:2012xq}
V.~F.~Cardone, N.~Radicella and S.~Camera,
Phys. Rev. D \textbf{85}, 124007 (2012)
[arXiv:1204.5294 [astro-ph.CO]].


 

\bibitem{Salako:2013gka}
I.~G.~Salako, M.~E.~Rodrigues, A.~V.~Kpadonou, M.~J.~S.~Houndjo and J.~Tossa,
JCAP \textbf{11}, 060 (2013)
[arXiv:1307.0730 [gr-qc]].




\bibitem{Nashed:2013bfa}
G.~G.~L.~Nashed,
Phys. Rev. D \textbf{88}, 104034 (2013)
[arXiv:1311.3131 [gr-qc]].


\bibitem{Ong:2013qja}
Y.~C.~Ong, K.~Izumi, J.~M.~Nester and P.~Chen,
Phys. Rev. D \textbf{88}, 024019 (2013)
[arXiv:1303.0993 [gr-qc]].

\bibitem{Otalora:2014aoa}
G.~Otalora,
Int. J. Mod. Phys. D \textbf{25}, no.02, 1650025 (2015)
[arXiv:1402.2256 [gr-qc]].


\bibitem{Farrugia:2016xcw}
G.~Farrugia, J.~Levi Said and M.~L.~Ruggiero,
Phys. Rev. D \textbf{93}, no.10, 104034 (2016)
[arXiv:1605.07614 [gr-qc]].
 
 
 
\bibitem{Cai:2018rzd}
Y.~F.~Cai, C.~Li, E.~N.~Saridakis and L.~Xue,
Phys. Rev. D \textbf{97}, no.10, 103513 (2018)
[arXiv:1801.05827 [gr-qc]].





 
\bibitem{Ferraro:2018tpu}
R.~Ferraro and M.~J.~Guzm{\'a}n,
Phys. Rev. D \textbf{97}, no.10, 104028 (2018)
[arXiv:1802.02130 [gr-qc]].
 



   
\bibitem{vandenHoogen:2023pjs}
R.~J.~van den Hoogen, A.~A.~Coley and D.~D.~McNutt,
JCAP \textbf{10}, 042 (2023)
[arXiv:2307.11475 [gr-qc]].
 
 

\bibitem{Jiang:2024otl}
X.~Jiang, X.~Ren, Z.~Li, Y.~F.~Cai and X.~Er,
Mon. Not. Roy. Astron. Soc. \textbf{528}, no.2, 1965-1978 (2024)
[arXiv:2401.05464 [gr-qc]].
 
 

 

 
\bibitem{Zhao:2024uzq}
J.~Y.~Zhao, M.~J.~Liu and K.~Yang,
Phys. Lett. B \textbf{860}, 139161 (2025)
[arXiv:2409.14471 [hep-th]].

 
 

  
  

\bibitem{Fenwick:2024jby}
J.~G.~Fenwick and M.~Ghezelbash,
Eur. Phys. J. C \textbf{84}, no.12, 1316 (2024)
[arXiv:2407.05172 [gr-qc]].
 
 
 
 
\bibitem{Jawad:2025mzv}
A.~Jawad, N.~Videla, A.~Malik Sultan, N.~Myrzakulov, A.~Aslam and S.~Shaymatov,
Nucl. Phys. B \textbf{1019}, 117101 (2025)
 
 


\bibitem{SwagatMishra:2025wut}
S.~Swagat Mishra, N.~S.~Kavya and P.~K.~Sahoo,
Phys. Lett. B \textbf{872}, 140036 (2026)

  
 
\bibitem{Landry:2025whg}
A.~Landry, Y.~Sekhmani, S.~K.~Maurya, A.~Ali and E.~N.~Saridakis,
[arXiv:2508.06290 [gr-qc]].

\bibitem{Manzoor:2026pyq}
R.~Manzoor, M.~Yousaf, Z.~Ikram and A.~Siddiqa,
Eur. Phys. J. C \textbf{86}, no.2, 193 (2026)

 
\bibitem{Bouhmadi-Lopez:2026dte}
M.~Bouhmadi-L{\'o}pez, C.~G.~Boiza, M.~Petronikolou and E.~N.~Saridakis,
Universe \textbf{12}, 81 (2026)
[arXiv:2601.22225 [gr-qc]].
 
\bibitem{Yang:2012hu}
J.~Yang, Y.~L.~Li, Y.~Zhong and Y.~Li,
Phys. Rev. D \textbf{85}, 084033 (2012)
[arXiv:1202.0129 [hep-th]].


 
\bibitem{Menezes:2014bta}
R.~Menezes,
Phys. Rev. D \textbf{89}, no.12, 125007 (2014)
[arXiv:1403.5587 [hep-th]].

 

 

\bibitem{Guo:2015qbt}
W.~D.~Guo, Q.~M.~Fu, Y.~P.~Zhang and Y.~X.~Liu,
Phys. Rev. D \textbf{93}, no.4, 044002 (2016)
[arXiv:1511.07143 [hep-th]].

\bibitem{Yang:2017evd}
K.~Yang, W.~D.~Guo, Z.~C.~Lin and Y.~X.~Liu,
Phys. Lett. B \textbf{782}, 170-175 (2018)
[arXiv:1709.01047 [hep-th]].



 






\bibitem{Moreira:2021xfe}
A.~R.~P.~Moreira, J.~E.~G.~Silva, F.~C.~E.~Lima and C.~A.~S.~Almeida,
Phys. Rev. D \textbf{103} (2021) no.6, 064046.

  


\bibitem{Moreira:2021vcf}
A.~R.~P.~Moreira, J.~E.~G.~Silva and C.~A.~S.~Almeida,
Eur. Phys. J. C \textbf{81}, no.4, 298 (2021)
[arXiv:2104.00195 [gr-qc]].



\bibitem{Guo:2018tpo}
W.~D.~Guo, Y.~Zhong, K.~Yang, T.~T.~Sui and Y.~X.~Liu,
Phys. Lett. B \textbf{800}, 135099 (2020)
[arXiv:1805.05650 [hep-th]].

 
 
\bibitem{Wang:2018jsw}
J.~Wang, W.~D.~Guo, Z.~C.~Lin and Y.~X.~Liu,
Phys. Rev. D \textbf{98}, no.8, 084046 (2018)
[arXiv:1808.00771 [hep-th]].


\bibitem{Moreira:2021uod}
A.~R.~P.~Moreira, F.~C.~E.~Lima, J.~E.~G.~Silva and C.~A.~S.~Almeida,
Eur. Phys. J. C \textbf{81} (2021) no.12, 1081.

 


\bibitem{Moreira:2023pes}
A.~R.~P.~Moreira, F.~M.~Belchior, R.~V.~Maluf and C.~A.~S.~Almeida,
Eur. Phys. J. C \textbf{83}, no.1, 48 (2023).


 

\bibitem{Moreira:2024zio}
A.~R.~P.~Moreira, S.~H.~Dong and E.~N.~Saridakis,
Class. Quant. Grav. \textbf{42}, no.7, 075013 (2025)
[arXiv:2407.15190 [gr-qc]].



 
\bibitem{Kofinas:2014owa}
G.~Kofinas and E.~N.~Saridakis,
Phys. Rev. D \textbf{90}, 084044 (2014)
[arXiv:1404.2249 [gr-qc]].


\bibitem{Kadam:2024rwd}
S.~A.~Kadam and B.~Mishra,
Phys. Dark Univ. \textbf{46} (2024), 101693
[arXiv:2401.15125 [gr-qc]].

\bibitem{Kadam:2022daz}
S.~A.~Kadam, B.~Mishra and J.~Levi Said,
Phys. Scripta \textbf{98} (2023) no.4, 045017
[arXiv:2210.06166 [gr-qc]].

\bibitem{Azhar:2020coz}
N.~Azhar, A.~Jawad and S.~Rani,
Phys. Dark Univ. \textbf{30}, 100724 (2020)
[arXiv:2009.13293 [gr-qc]].




\bibitem{Sharif:2018sgg}
M.~Sharif and K.~Nazir,
Annals Phys. \textbf{393} (2018), 145-166
[arXiv:1805.03528 [gr-qc]].

\bibitem{Farrugia:2018gyz}
G.~Farrugia, J.~Levi Said, V.~Gakis and E.~N.~Saridakis,
Phys. Rev. D \textbf{97} (2018) no.12, 124064
[arXiv:1804.07365 [gr-qc]].


\bibitem{delaCruz-Dombriz:2017lvj}
{\'A}.~de la Cruz-Dombriz, G.~Farrugia, J.~L.~Said and D.~Saez-Gomez,
Class. Quant. Grav. \textbf{34} (2017) no.23, 235011
[arXiv:1705.03867 [gr-qc]].






\bibitem{Ilyas:2025pvh}
M.~Ilyas, Z.~Yousaf and M.~Z.~Bhatti,
Nucl. Phys. B \textbf{1020} (2025), 117161.

\bibitem{Rehman:2025ikt}
A.~Rehman, M.~Yousaf, F.~Javed and P.~Channuie,
Eur. Phys. J. C \textbf{85} (2025) no.9, 949.

\bibitem{Moreira:2025vsc}
A.~R.~P.~Moreira and S.~H.~Dong,
Nucl. Phys. B \textbf{1018} (2025), 117025.

\bibitem{Ilyas:2025fal}
M.~Ilyas and Z.~U.~N.~Shinwari,
JHEAp \textbf{48} (2025), 100414.

\bibitem{Samaddar:2025dyj}
A.~Samaddar and S.~S.~Singh,
Phys. Dark Univ. \textbf{50}, 102081 (2025)
[arXiv:2506.06388 [gr-qc]].





\bibitem{Akbarieh:2025vau}
A.~R.~Akbarieh, N.~S.~Ilkhchi and Y.~Heydarzade,
Phys. Lett. B \textbf{875} (2026), 140345
[arXiv:2506.04469 [gr-qc]].

\bibitem{Dimakis:2025wtc}
N.~Dimakis, A.~Giacomini, G.~Leon, A.~Paliathanasis, E.~Pozdeeva and S.~Vernov,
Gen. Rel. Grav. \textbf{57} (2025) no.10, 153
[arXiv:2505.02663 [gr-qc]].

\bibitem{Ilyas:2026qbi}
M.~Ilyas, K.~Masood and N.~A.~Shah,
Annals Phys. \textbf{486} (2026), 170334.

\bibitem{Dubey:2025nuf}
V.~C.~Dubey,
Eur. Phys. J. C \textbf{85} (2025) no.12, 1399.





 \end{thebibliography}
\end{document}